%% file: main_v2.tex
\documentclass[
    twocolumn,
	prd,
	amssymb,
	preprintnumbers,superscriptaddress,
	nofootinbib]{revtex4-1}

\pdfoutput=1
\usepackage{setspace}
\usepackage{paralist}
\usepackage{graphicx}
\usepackage{enumitem}
\usepackage{latexsym}
\usepackage{amsfonts}
\usepackage{xcolor}
\usepackage[export]{adjustbox}
\usepackage{amssymb}
\usepackage{amsmath}
\usepackage[thinlines]{easytable}
\usepackage{slashed}
\usepackage{dcolumn}
\usepackage{verbatim}
\usepackage{float}
\usepackage{multirow}
\usepackage{xspace}
\usepackage[normalem]{ulem}
\usepackage[
pdfauthor={Jeremy Sakstein}]{hyperref}
\usepackage{tabularx}

\input{universalnewcommands.tex}

\newcommand{\GeV}{\text{GeV}}

\newcommand{\msun}{{\rm M}_\odot}

\newcommand{\mbh}{M_{\rm BH}}

\DeclareRobustCommand{\okina}{%
  \raisebox{\dimexpr\fontcharht\font`A-\height}{%
    \scalebox{0.8}{`}%
  }%
}

\allowdisplaybreaks

\begin{document}

\title{Light Axion Emission and the Formation of Merging Binary Black Holes}
\author{Djuna Croon} \email{djuna.l.croon@durham.ac.uk}
\affiliation{Institute for Particle Physics Phenomenology, Department of Physics, Durham University, Durham DH1 3LE, U.K.}
\author{Jeremy Sakstein} \email{sakstein@hawaii.edu}
\affiliation{Department of Physics \& Astronomy, University of Hawai\okina i, Watanabe Hall, 2505 Correa Road, Honolulu, HI, 96822, USA}

\date{\today}

\begin{abstract}
We study the impact of stellar cooling due to light axion emission on the formation and evolution of black hole binaries, via stable mass transfer and the common envelope scenario.~We find that in the presence of light axion emission, no binary black hole mergers are formed with black holes in the lower mass gap ($\mbh < 4 \msun $) via the common envelope  formation channel.~In some systems, this happens because axions prevent Roche lobe overflow.~In others, they prevent the common envelope from being ejected.~Our results  apply to axions with couplings $ g_{a \gamma} \gtrsim 10^{-10}\, \rm GeV^{-1}$ (to photons) or $\alpha_{ae} \gtrsim 10^{-26} $ (to electrons) and masses $ m_a \ll 10 \, \rm keV$.~Light, weakly coupled particles may therefore apparently produce a mass gap $2 \msun <  \mbh < 4 \msun  $ in the LIGO/Virgo/KAGRA data, when no mass gap is present in the stellar remnant population.
\end{abstract}

\preprint{IPPP/22/59}

\maketitle

%%%%%%%%%%%%%%%%%%%%%%%%%
\section{Introduction}
%%%%%%%%%%%%%%%%%%%%%%%%%
The detection of gravitational waves (GW) emitted by binary black holes (BBHs) is revolutionizing our understanding of black hole properties, compact object formation, and stellar evolution.~The LIGO/Virgo/KAGRA (LVK) collaboration has to date observed $\sim 90$ BBH events \cite{LIGOScientific:2021djp}, from which conclusions can be drawn affecting not just astrophysics \cite{LIGOScientific:2021psn}, but also gravitation \cite{Straight:2020zke}, and nuclear and particle physics (e.g.~ \cite{Farmer:2020xne,Croon:2020oga,Sakstein:2020axg,Baxter:2021swn}).

In this work, we study the effect of novel particles on the formation and evolution of BBH binaries.~
A very important formation mechanism is the common-envelope (CE) scenario (see, e.g.~\cite{paczynski1976structure,van1976structure,tutukov1993merger,hurley2002evolution,Ivanova:2011tc,ivanova2013common,mapelli2020binary}).~In this scenario, post-main sequence (MS) stars in close binaries undergo Roche lobe overflow (RLOF) and form a CE.~The strong gas drag from the envelope causes the stars to lose kinetic energy and inspiral.~Crucially, the CE may be ejected due to the transfer of thermal energy, leaving a close binary of a black hole and the core of a giant star.~The star may then collapse into a black hole; if this happens without a strong natal kick, the resulting system is a BBH which may merge within a Hubble time $t_h$.~ 

The existence of a lower mass gap --- between the heaviest neutron stars and the lightest black holes --- is currently uncertain.~Mass measurements in X-ray transients have identified black holes with possible masses as low as $ 2.1 \msun$ (in the case of GRO J0422+32) \cite{Kreidberg:2012ud}.~The GWTC-3 catalogue includes events with secondary objects within the mass gap, such as GW190814 \cite{LIGOScientific:2020zkf}.~Thorough theoretical investigations of binaries containing light black holes are needed, along with informed gravitational wave data analysis, to conclusively determine the existence and location of the lower mass gap.

As we will show, light axion ($m_a \ll 10$ keV) emission affects  stellar binaries in such a way that CE ejection does not occur for systems where the secondary object is a BH within the lower mass gap.
We model our binaries as combinations of a post-MS supergiant and a lighter black hole and simulate the evolution until either the system merges due to angular momentum loss via stable mass transfer (SMT) or CE evolution, or the supergiant reaches the end of core carbon burning.~We observe two important effects of axion emission:
\begin{enumerate}
    \item In some systems, the axion emission prevents RLOF so that the objects never interact.~
    \item In other systems, axion emission prevents the CE from being ejected and the objects merge {before the primary collapses into a black hole}.
\end{enumerate}
In both scenarios, no BBH systems are formed in the lower mass gap.~Thus, axions may cause a lower mass gap for black holes in \emph{black hole binaries} even if no such mass gap exists for isolated black holes.~Specifically, our simulations predict no BBH systems with BH masses $2\msun\le \mbh\le 4\msun$ {formed via the CE scenario}.~

This paper is organized as follows.~In section \ref{sec:BBH} we discuss the formation channels for binary black hole formation and present the details of how stable mass transfer and the common envelope phase are implemented into the stellar structure code we use.~In section \ref{sec:emission} we describe the effects of new particle emission on stellar structure, and introduce the specific axion production processes we study in this work.~
We present our results in section \ref{sec:results}.~These are discussed in section \ref{sec:conclusions} where we also conclude.

%%%%%%%%%%%%%%%%%%%%%%%%%
\section{Black hole binaries}
\label{sec:BBH}
%%%%%%%%%%%%%%%%%%%%%%%%%
\subsection{Black hole binary formation}

Inspiraling BBHs undergo several stages of evolution.~
The gravitational waves from these systems observed by the LVK collaboration probe the very last stage of the binary's evolution, as this emission only starts to dominate the energy loss at very small radii.~While these observations provide a wealth of information about the properties of the final moments of the binary's life, less is known about the preceding stages.~The current gravitational wave catalogue shows some evidence pointing towards multiple formation channels \cite{Zevin:2020gbd,LIGOScientific:2021psn}.~Forecasting merger rates detectable via GWs and interpreting the GW data requires detailed studies of binary formation and evolution.~

Several proposed BBH formation mechanisms exist.~Of particular interest for stellar evolution are isolated binaries, as the processes in the star can in principle be studied without detailed knowledge of the dynamics of the population.~Isolated stellar binaries can lose enough energy without external effects due to gas drag in CE-evolution.~In the absence of the CE, such stellar binaries are not expected to merge within a Hubble time.~
Crucial to the formation of a BBH system, the CE needs to be ejected before the merger occurs.~

Alternative BBH formation mechanisms have been proposed, including stable mass transfer \cite{van2017forming,neijssel2019effect}, over-contact binary evolution in which both stars in the binary have large spin, which prevents single premature black hole formation \cite{Marchant:2016wow}, and chemically homogeneous evolution \cite{1987A&A...178..159M,mandel2016merging} in which an initially compact binary can remain so due to rotationally-induced mixing that prevents the envelope from expanding.~If the binary is not formed in isolation, three-body encounters can be an important formation mechanism \cite{thompson2011accelerating,antonini2017binary,Vigna-Gomez:2020fvw}.~In this work, we will focus on the CE mechanism, but the effects of new particle emission on these alternative mechanisms would be interesting topic for future studies.~

%%%%%%%%%%%%%%%%%%%%%%%%%
\subsection{Modelling binary evolution}\label{sec:modelling}
%%%%%%%%%%%%%%%%%%%%%%%%%

In this work we focus on BBH formation in isolated stellar systems via the CE scenario.~
We simulate the evolution of a $ 30\msun$ star with metallicity $Z=0.02$ in a circular orbit with a lighter BH companion with masses in the range $M_{\rm BH} = 0.5-5 \msun$.~The BH is modeled as a point mass.~The evolution of these binaries is simulated for different initial periods in the range $P_i = 1000-2000$ days.\footnote{Simulations with $ P_i \leq 1000$ days and $ 5 \msun \leq M_{\rm BH} \leq 18 \msun$ yielded no differences in outcome due to axion emission.} We use the stellar structure code MESA version 15140 \cite{Paxton:2010ji,Paxton:2013pj,Paxton:2015jva,Paxton:2017eie,Paxton:2017eie}.~MESA is a one-dimensional code but is capable of simultaneously evolving binary stars and their orbital dynamics using a series of approximations described in \cite{Paxton:2015jva}.~MESA can account for stable mass transfer through the L1 Lagrangian point as well as outflows from L2 and L3.

In addition to stable mass transfer, MESA can also model unstable mass transfer via a common envelope.~We refer the interested to \cite{Paxton:2015jva,Marchant:2021hiv} for a full description of the schemes MESA uses to simulate stable/unstable mass transfer.~Here, we review only the salient features.~A description of the stellar physics used in our simulations is given in Appendix \ref{appendix:MESA}.~ A reproduction package containing the inlists, run\_star\_extras, and run\_binary\_extras used for our simulations is available here: \href{https://zenodo.org/record/6949679}{https://zenodo.org/record/6949679}.

If, during a phase of stable mass transfer, the mass loss rate exceeds a threshold $\dot{M}_{\rm high}$ then the evolution of the system proceeds through a CE phase.~During this phase, the binary loses orbital angular momentum and inspirals.~At any given timestep, MESA calculates the decreases in the orbital separation $a$ (assuming the orbit remains circular) by equating the binding energy of the ejected layers $E_{\rm bind}$ to the orbital energy $E_{\rm orb}$ i.e.,
\begin{eqnarray}
\label{eq:CE_eject}
     E_{\rm bind} = \alpha_{\rm CE} E_{\rm orb},
\end{eqnarray}
where $\alpha_{\rm CE}$ is a free parameter that describes the CE ejection efficiency.~The binding energy is calculated as
\begin{eqnarray}
\label{eq:binding_energy}
    E_{\rm bind} = \int_{M_{\rm core}}^{M_{d, i}}\left(-\frac{GM}{r}+\alpha_{\rm th} u\right)\mathrm{d} m,
\end{eqnarray}
where `d' refers to the donor, `i' refers to the pre-CE mass, $u$ is the specific internal energy of the gas (which includes contributions from hydrogen and helium recombination), and $\alpha_{\rm th}$ is a free parameter that describes the efficiency with which thermal energy can be used to eject the envelope \cite{1995MNRAS.272..800H}.~The orbital energy is
\begin{eqnarray}
    \label{eq:orbital_energy}
    \Delta E_{\rm orb} = -\frac{G M_{d,f}M_{ac,f}}{2a_f} + \frac{G M_{d,i}M_{ac,i}}{2a_i},
\end{eqnarray}
where $G$ is Newton's constant, $a$ is the orbital separation, `d' refers to the donor, `ac' refers to the accretor, `i' refers to the pre-CE properties, and `f' refers to the post-CE properties.~MESA solves equation \ref{eq:CE_eject} for $a_f$ to determine the post-CE orbital separation.~The final masses are computed using the descriptions described below.~In this work, we take $\alpha_{\rm CE}=\alpha_{\rm th}=1$.~We have verified that these choices do not alter our conclusions.

During the CE phase, the donor star loses mass.~This is modeled as follows.~When the star's radius $R$ is larger than the Roche Lobe (RL), $R_{\rm RL}$, the mass loss is given by a constant rate $\dot{M}_{\rm high}$.~As the star loses mass, it may eject the CE, in which case its radius will begin to recede inside the Roche Lobe.~The point of detachment is defined as $R/R_{\rm RL}<1-\delta$ ($\delta$ is a free parameter) at which point the CE phase ends.~If this criterion has not been reached before the merger or before carbon depletion, we consider the merger to be a CE merger.~When $ 1-\delta < R/R_{\rm RL} < 1$ the mass loss rate is reduced to \cite{Marchant:2021hiv}
\begin{align}
\label{eq:masslossratelow}
    \log_{10}(\dot{M}_{\rm CE})&= \log_{10}(\dot{M}_{\rm high}) \nonumber\\&+ \frac{1-R/R_{\rm RL}}{\delta}\log_{10}\left(\frac{\dot{M}_{\rm low}}{\dot{M}_{\rm high}}\right).
\end{align}
In this work we take $\dot{M}_{\rm high}=1\msun/\textrm{yr}$, $\dot{M}_{\rm low}=10^{-5}\msun/\textrm{yr}$ and $\delta=0.02$.~We have tested explicitly that our results are insensitive to these choices.

If the system is able to eject its common envelope or the mass loss proceeds solely via stable mass transfer the star will ultimately collapse to form a BH of identical mass since the pair-instability is not encountered for the $30\msun$ model studied in this work \cite{Marchant:2018kun,Farmer:2019jed,Croon:2020ehi,Croon:2020oga}.~The resultant BBH system will merge in a time (assuming a circular orbit) \cite{1964PhRv..136.1224P}
\begin{equation}
\label{eq:peters_time}
    t_{\rm m} = \frac{5c^5}{256G^3}\frac{(1+q)^2}{qM^3}a^4
\end{equation}
where $q=m_2/m_1$ and $M=m_1+m_2$ with $m_1$ and $m_2$ the primary and secondary masses respectively\footnote{Equation \eqref{eq:peters_time} is valid to leading-order in the post-Newtonian expansion and therefore strictly describes the inspiral phase.~The timescale for the merger phase is negligible compared with the inspiral phase so we take the merger time to be equal to the inspiral time.}.~Only systems that merge within a Hubble time will be detected by LVK.~We take the Hubble constant to be $H_0=70$ km/s/Mpc in this work.

\begin{figure}
    \centering
    \includegraphics[width=0.5\textwidth]{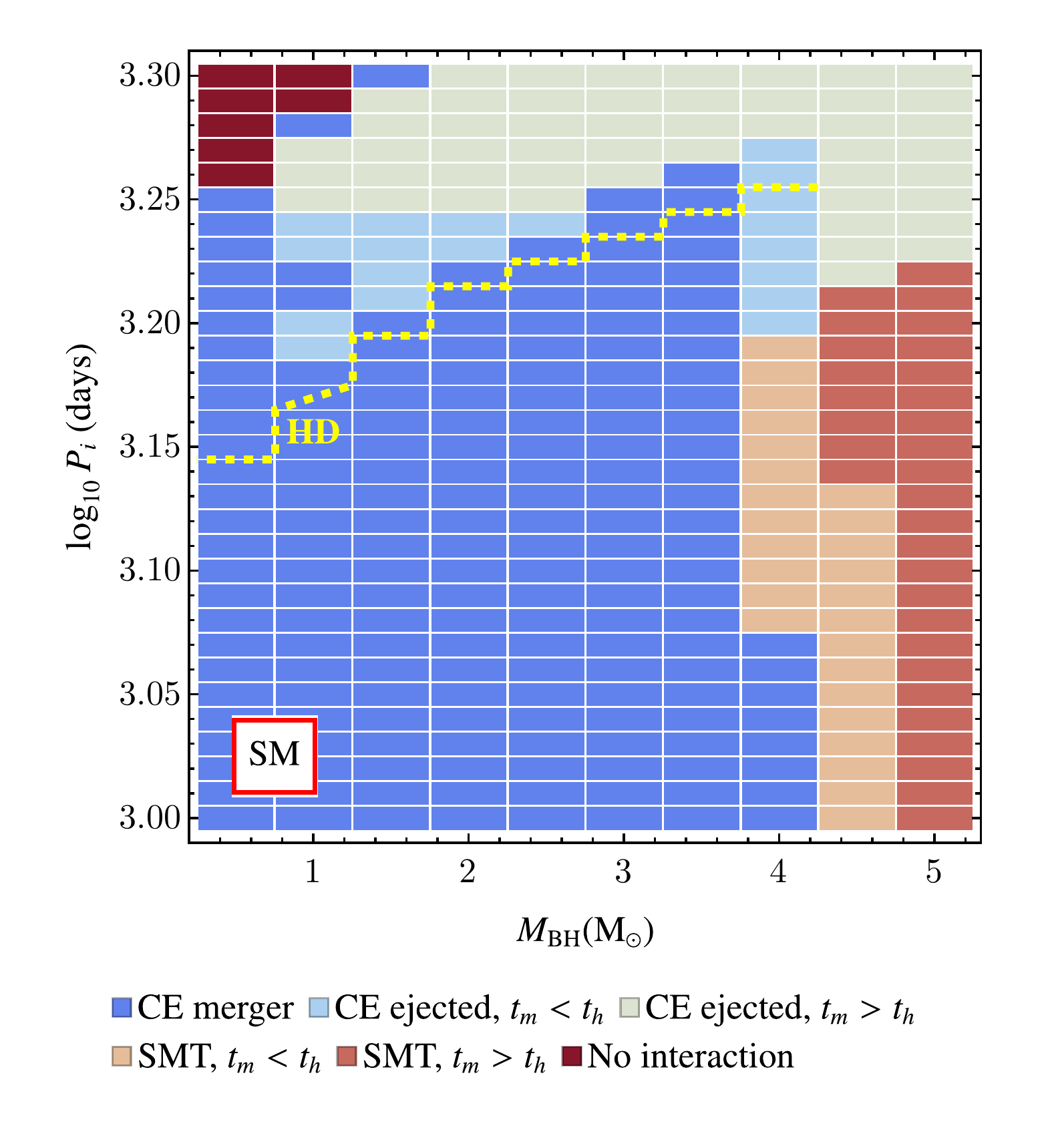}
    \caption{Merger outcomes in the SM.~The dashed yellow line divides CE scenarios for which the CE switches on after helium depletion (above the line) and before helium depletion (note that this is a different definition than used in \cite{Marchant:2021hiv}).}
    \label{fig:SM_outcomes}
\end{figure}

We ran a grid of simulations without new losses.~The parameters varied over the ranges $0.5\msun\le M\le 5\msun$ with $\Delta M=0.5\msun$ and $3.0\le\log_{10}(P/\textrm{days})\le3.3$ with $\Delta \log_{10}(P/\textrm{days})=0.01$.~The results are shown in Fig.~\ref{fig:SM_outcomes}.~These outcomes are consistent with those reported in \cite{Marchant:2021hiv}.~In particular, CE ejection happens in two specific regions, distinguished by the onset of CE evolution before and after helium depletion.~

\section{Light particle emission}
\label{sec:emission}
%%%%%%%%%%%%%%%%%%%%%%%%%
\subsection{Stellar response to light particle emission}
\label{sec:stellar response}
%%%%%%%%%%%%%%%%%%%%%%%%%

New light particles weakly coupled to the Standard Model are produced in the cores of stars.~They subsequently free-steam out of the star and act as a novel source of energy loss in addition to neutrino losses.~In this section, we briefly review the consequences of such novel loss channels at different stages in stellar evolution.

\begin{figure*}[t]
    \centering
    \vspace{-1cm}
    \includegraphics[width=\textwidth]{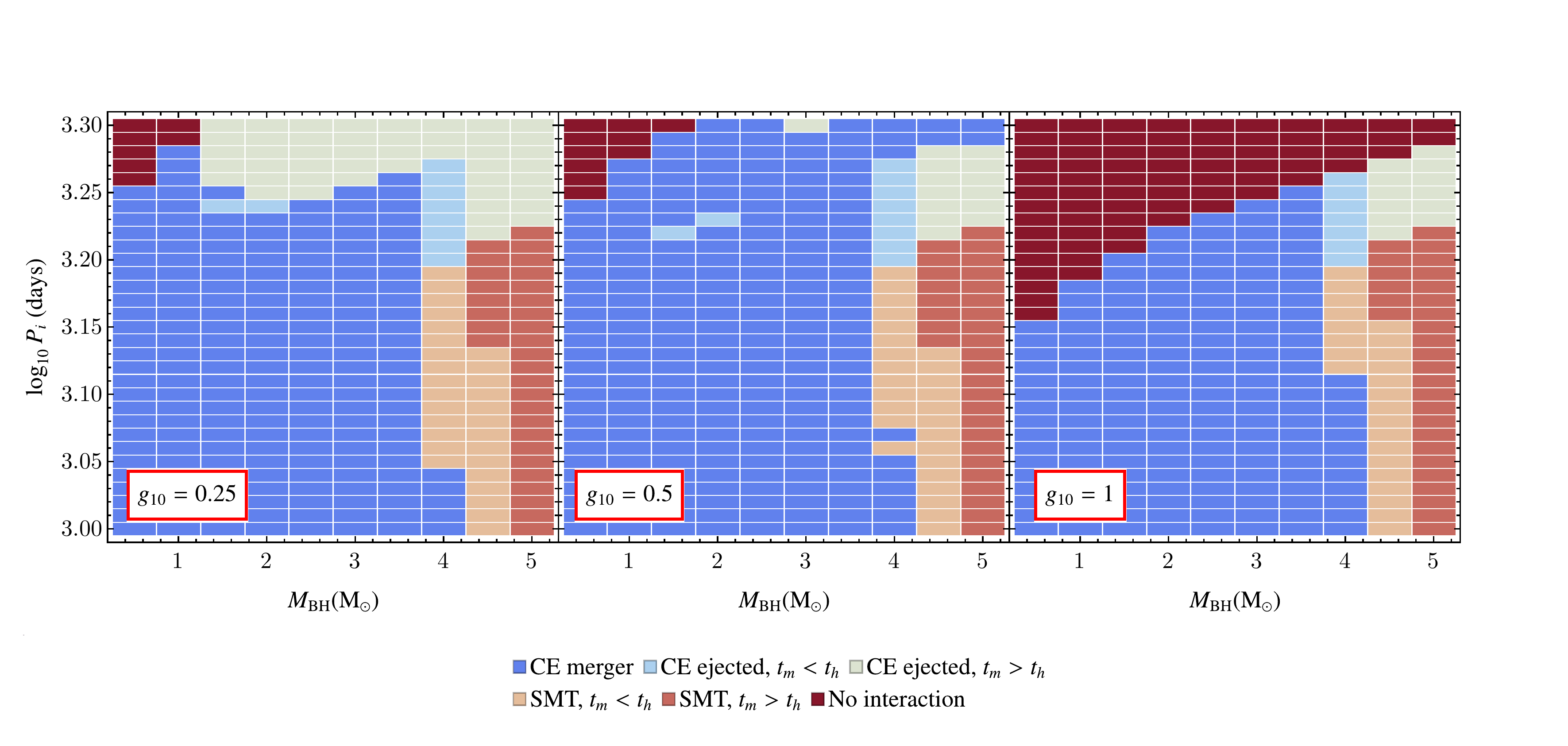}
    \vspace{-1cm}
    \caption{Outcomes of binary scenarios for axion emission through the axion-photon coupling $ g_{10}$ as a function of accretor mass $ \mbh$ and initial period $P_i$.~It is seen that for significant $g_{10}$, low mass companions only give rise to no interaction or mergers which take place during the CE phase.
    }
    \label{fig:g10_outcomes}
\end{figure*}

Under the assumption of a homologous transformation $r' = yr $, such that the entire profile of the star (including its density, radius, and temperature) can be rescaled by a common factor $y$, the results of new losses on a lower mass MS star were found to be contraction, heating, and luminosity increase \cite{frieman1987axions,raffelt1996stars} --- in other words, its evolution speeds up.~We are interested in higher mass stars and post-MS evolution, so we will generalise this treatment.

Assuming a chemically homogeneous star, we must have energy generation rate (per unit mass) $\epsilon$ and opacity $\kappa$ scaling with temperature $T$ and density $\rho$
\begin{equation}
    \epsilon \propto \rho^n T^\nu, \quad \quad \kappa \propto \rho^s T^p.
\end{equation}
From the equation for radiative transfer, and $ T' = y^{-1}T$, $ \rho' =y^{-3} \rho$, it then follows that the local energy flux scales as\footnote{An alternative form of this equation can be derived for convective regions
bordered by a thin photosphere (such that constant opacity can be assumed), $ L'(r') = y^{5/2} L(r)$ \cite{frieman1987axions}, and the following holds with the replacement  $3s+p \to 5/2 $.~See also \cite{Peled:2022byr}.}
\begin{equation}
    L'(r') = y^{3s+p} L(r).
\end{equation}
The new losses modify the energy generation rate $\epsilon$, defined as
\begin{equation}
\begin{split}
    \epsilon &= \epsilon_{\rm nuc} - \epsilon_{\rm grav} - \epsilon_{\rm neutrino} - \epsilon_{\chi}
    \\ &\equiv (1 - \delta_{\rm grav } - \delta_{\rm neutrino} - \delta_{\chi}) \epsilon_{\rm nuc}
\end{split}
\end{equation}
where $\epsilon_\chi$ are the new particle losses.~
From the energy generation equation ${\rm d} L / {\rm d}r = 4 \pi r^2 \epsilon \rho $,
\begin{equation}
    \begin{split}
        L'(r') &= y^{-(3 n + \nu)} (1 - \delta_{\rm grav } - \delta_{\rm neutrino} - \delta_{\chi}) L(r)
    \end{split}
\end{equation}
we can conclude 
\begin{equation}
     y = (1 - \delta_{\rm grav } - \delta_{\rm neutrino} - \delta_{\chi})^{\frac{1}{3s + p +3n +\nu}},
\end{equation}
such that for small $\sum\delta \equiv \delta_{\rm grav } + \delta_{\rm neutrino} + \delta_{\chi}$, 
\begin{equation}
\begin{split}
    \frac{\delta R}{R} &= \frac{-\sum\delta}{3s + p +3n +\nu} 
    \\ \frac{\delta L}{L} &=
    \frac{- (3s+p) \sum\delta }{3s + p +3n +\nu}
    \\ \frac{\delta T}{T} &=
    \frac{\sum\delta}{3s + p +3n +\nu}
\end{split}
\label{eq:homologyscaling}
\end{equation}
We may assume $ s= 0$ and $p =0 $ for the post-MS evolution of a high mass star\footnote{For a low-mass MS star with $ T < 10^7$ K, Kramers' opacity law gives $p=-7/2$ and $s=1$, but for higher temperatures electron scattering drives $\kappa $ to a constant.~
}
At the temperatures we are interested in, $\nu \sim 17-40$, and $ n = 1-2$, where the lower numbers hold for the CNO cycle and the larger numbers for the triple-$\alpha$ process.
Using these values in \eqref{eq:homologyscaling} implies that in the presence of new losses, the radii of large post-MS stars decreases, the temperature increases, and the luminosty is not affected (this is a result of the constant opacity).
While this approximation is not in general valid because the new losses scale differently with temperature and density than the nuclear rates, we expect it to be a first approximation when interpreting the new losses as an average over the star.~

%%%%%%%%%%%%%%%%%%%%%%%%%
\subsection{Axion emission rates}
%%%%%%%%%%%%%%%%%%%%%%%%%
In what follows, we will consider axions coupled to photons and electrons via the following Lagrangian:
\begin{equation}
    \label{eq:Laxion}
    \cL_{\rm axion} =  - \frac14 g_{a\gamma} a F_{\mu \nu} \widetilde F^{\mu\nu}  - i g_{ae} a \bar\psi_e \gamma_5 \psi_e,  
\end{equation}
where we have neglected the axion mass since we are interested in the regime $m_a\ll10\textrm{ keV}$ and couplings to other particles not relevant for this study.~

The axion-photon coupling parameterized by $g_{a\gamma}$ gives rise to axion production in stars via the Primakoff process as well as others that are highly subdominant at the low masses considered here e.g., photon coalescence \cite{DiLella:2000dn,Carenza:2020zil,Lucente:2022wai}.~The Primakoff energy loss rate per unit mass is \cite{Raffelt:1990yz, Choplin:2017auq,Choplin:2017auq}
\begin{equation}
   \mathcal{Q}_{a\gamma}= 283.16g_{10}^2T_8^7\rho_3^{-1}g(\xi^2)\textrm{ ergs/g/s},
\end{equation}
where $T_8\equiv T/10^8$K, $\rho_3\equiv\rho/10^3\textrm{ gcm}^{-3}$, $g_{10} \equiv g_{a\gamma}/(10^{10}\GeV^{-1})$, and $\xi = k_S/2T$
with $k_S$ the Debye momentum
\begin{equation}
\label{eq:debyeMomentum}
    k_S^2=\frac{4\pi\alpha}{T}\sum_{i} n_iZ_i^2,
\end{equation}
where the sum runs over both ions and electrons.~The function $g(\xi^2)$ is well-approximated by \cite{Friedland:2012hj}
\begin{equation}
\scalebox{.8}{$\displaystyle
    g(\xi^2)=\left(\frac{1.037 \xi^2}{1.01+{\xi^2}/{5.4}}+\frac{1.037 \xi^2}{44+0.628 \xi^2}\right) \log \left(3.85\, +\frac{3.99}{\xi^2}\right).$}
\end{equation}

\begin{figure*}[t]
    \centering
    \vspace{-1cm}
    \includegraphics[width=\textwidth]{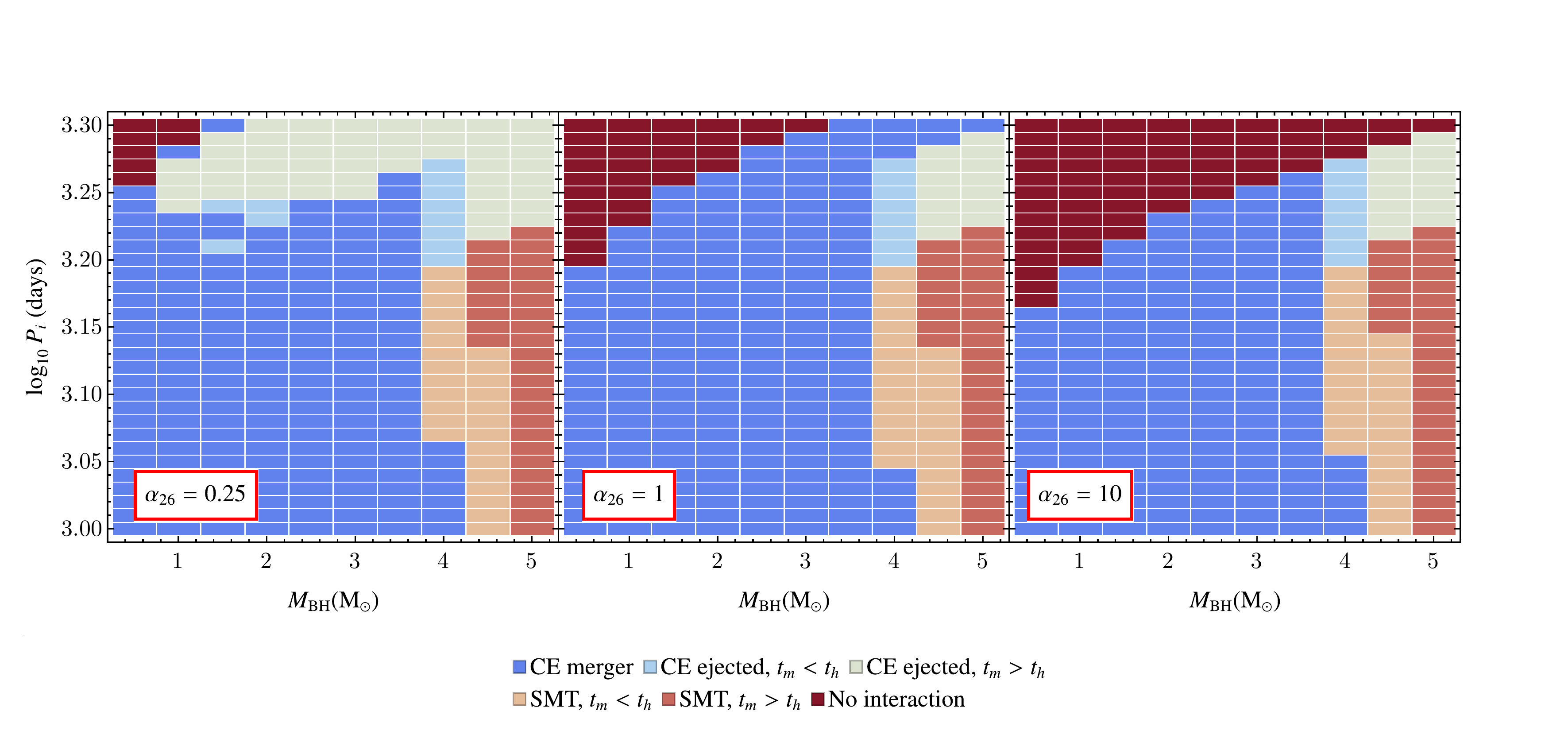}
    \vspace{-1cm}
    \caption{
    Same as figure \ref{fig:g10_outcomes} but for the axion-electron coupling $\alpha_{26}$.~}
    \label{fig:ae_outcomes}
\end{figure*}

The axion-electron coupling gives rise to axion production via bremsstrahlung emission, atomic processes (axio-recombination and atomic de-excitation), and semi-Compton scattering.~Bremstrahlung emission is negligible compared with the latter at the core temperatures and densities relevant for this study so we have neglected it in our simulations. Appendix \ref{app:Sevol} shows typical values of some stellar  quantities, including the Debye screening length, for the stars studied in this work.~Our reproduction package \cite{djuna_croon_2022_6949679} includes the option of including  bremsstrahlung processes so we have provided a description of these in Appendix \ref{app:Brem}.~We assume that the axion production rate per unit mass is dominated by semi-Compton scattering $e+\gamma\rightarrow e+a$, which is given by \cite{Raffelt:1990yz,Raffelt:1994ry}
\begin{equation}
    \mathcal{Q}_{\rm sC} = 33\alpha_{26}Y_eT_8^6F_{\rm deg}\textrm{ ergs/g/s},
\end{equation}
where $\alpha_{26}\equiv10^{26} g^2_{ae}/4\pi$, $Y_e$ is the number of electrons per baryon, and $F_{\rm deg}$ encodes the effects of Pauli blocking due to electron degeneracy.~$F_{\rm deg}$ is well approximated by \cite{Croon:2020ehi,Croon:2020oga}
\begin{align}
  F_{\rm deg} &= \frac{1}{2} \left[ 1-\tanh  f(\rho,T) \right] \\ f(\rho,T) &=
    a \log_{10} \left[ \frac{\rho}{\text{g cm}^{-3}}\right] -b \log_{10}\left[ \frac{T}{\rm K}\right] +c,
\end{align}
with $a=0.973$, $b=1.596$, and $c=8.095$.~$F_{\rm deg}\approx 1$ for the ranger of temperatures and densities relevant for this work.~

%%%%%%%%%%%%%%%%%%%%%%%%%
%\section{Binary effects of light particle emission}
\section{Results}
\label{sec:results}
%%%%%%%%%%%%%%%%%%%%%%%%%

We show the results of our simulations in Figs.~\ref{fig:g10_outcomes} and~\ref{fig:ae_outcomes} for the axion-photon coupling $g_{10}$ and axion-electron coupling $\alpha_{26}$ respectively (with the other coupling set to zero) respectively.~The parameters were varied over the ranges $0.5\msun\le M\le 5\msun$ with $\Delta M=0.5\msun$ and $3.0\le\log_{10}(P_i/\textrm{days})\le3.3$ with $\Delta \log_{10}(P_i/\textrm{days})=0.01$.~The plots show the outcomes of the simulations, as described in section~\ref{sec:modelling}.

The first important observation is that axion emission before CE evolution leads to contraction of the radius, consistent with equation \eqref{eq:homologyscaling}.~As a result, the models in the top left corner of Figs.~\ref{fig:g10_outcomes} and~\ref{fig:ae_outcomes} never undergo RLOF.~These models reach carbon depletion without any interaction.~This primarily affects binaries with lower companion masses because the radius of the Roche Lobe is a decreasing function of accretor mass.~As can be seen, the greater the axion coupling, the larger the region in both companion mass and initial binary period this applies to.~

The second important result is that axion emission implies models which undergo CE evolution after helium depletion are less likely to eject the CE, as can be identified from the shrinking region of CE-ejected mergers for low mass companions and large initial periods in Figs.~\ref{fig:g10_outcomes} and \ref{fig:ae_outcomes}.
We find that this outcome can be the result of two situations.~The star either (1) never recedes into its Roche Lobe, and loses so much mass during the CE phase that the CE is removed; or (2) does not recede quickly enough into its Roche Lobe to eject its envelope before it is removed.~Both of these imply a CE merger takes place where otherwise CE ejection would have.~This effect too is enhanced for larger couplings.~

We find that axion emission does not significantly affect the outcome of the CE phase if it occurs before helium depletion.~Axion losses can significantly alter the evolution during and after He-burning, but not during the main-sequence evolution, since the temperatures are too low ($T_8<1$) for efficient axion production.~Moreover, whether CE onset occurs before or after core helium depletion changes the duration of the CE phase.~In the former case, the CE phase lasts $\leq \mathcal{O}(10^2) $ years in most cases, whereas in the latter case the CE phase can take up to $\mathcal{O} (10^4)$ years, giving more time for it to be effected by axion emission.~

\begin{figure}[b]
    \centering
    \includegraphics[width=.5\textwidth]{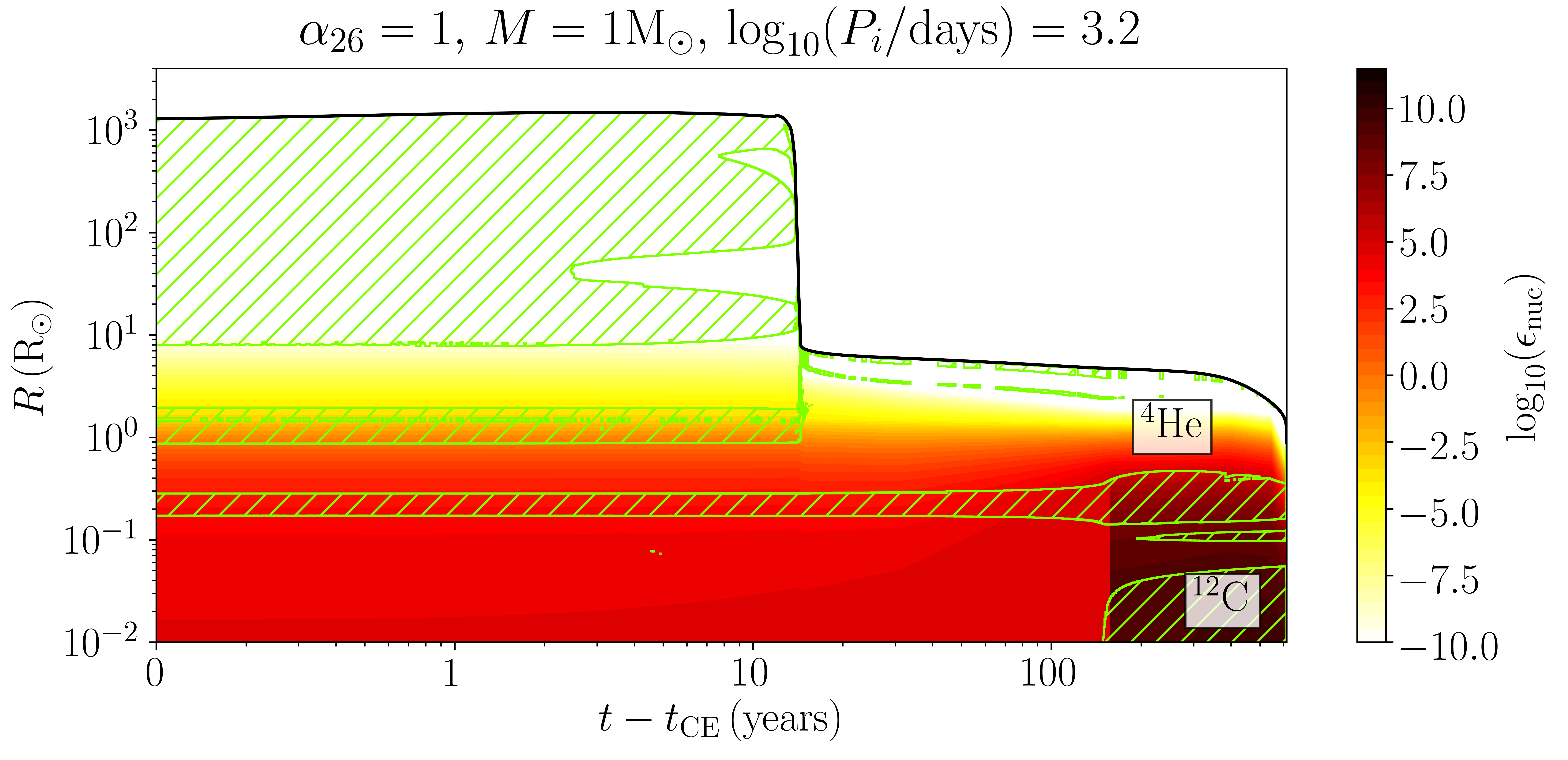}

    \includegraphics[width=.5\textwidth]{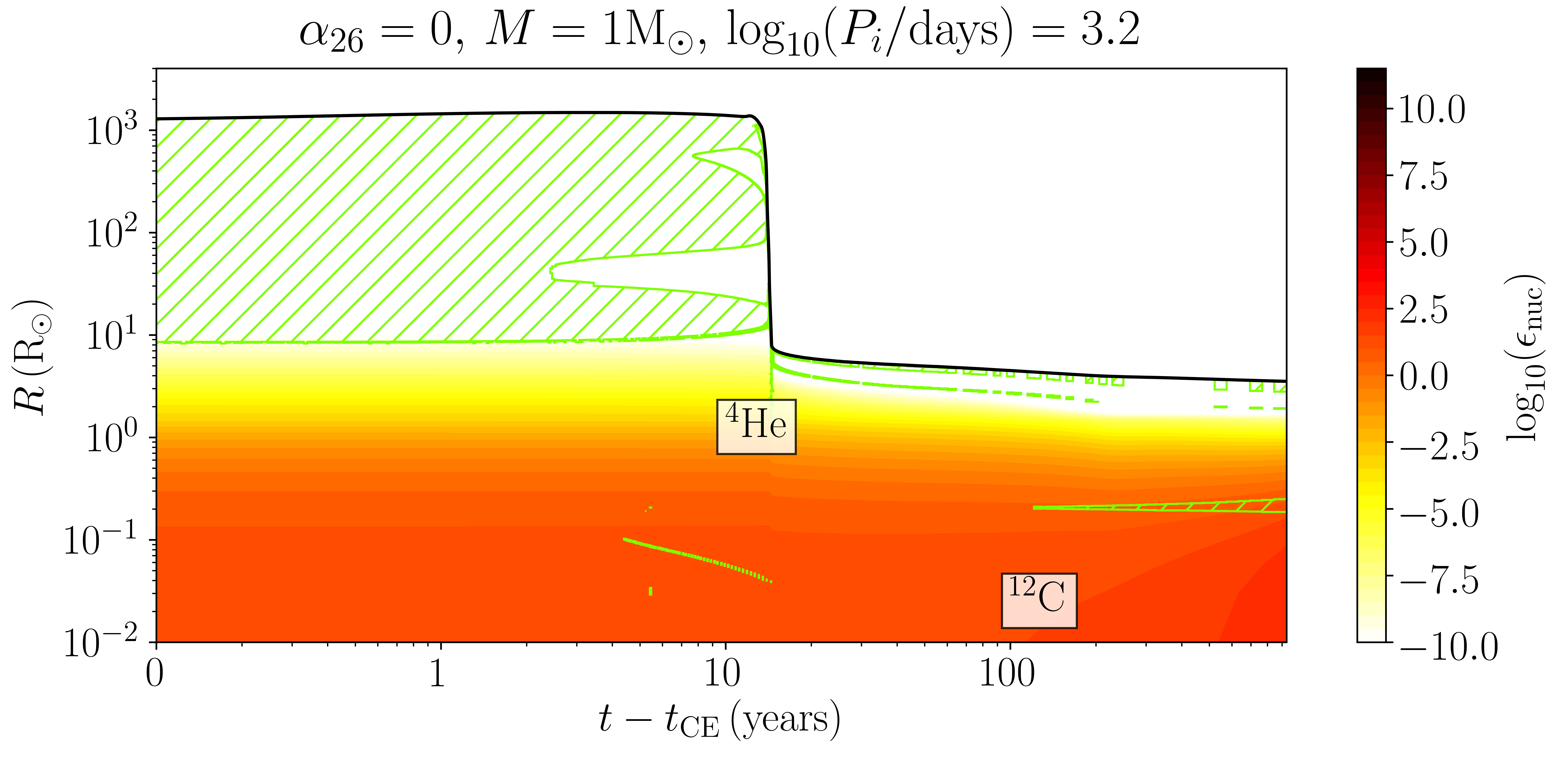}
    \caption{Nuclear burning rates for binaries with companion mass $1 \msun$ and initial period $\log_{10}(P_i/\textrm{days})=3.2$ for the case $ \alpha_{26} = 1$ (\emph{upper panel}) and the SM (\emph{lower panel}).
    In the upper panel, the right edge corresponds to the merger; in the lower panel, the right edge corresponds to CE ejection.~ Convective regions are shown using green hatching.
    }
    \label{fig:nucburning}
\end{figure}

\begin{figure}
    \centering
    \includegraphics[width=.5\textwidth]{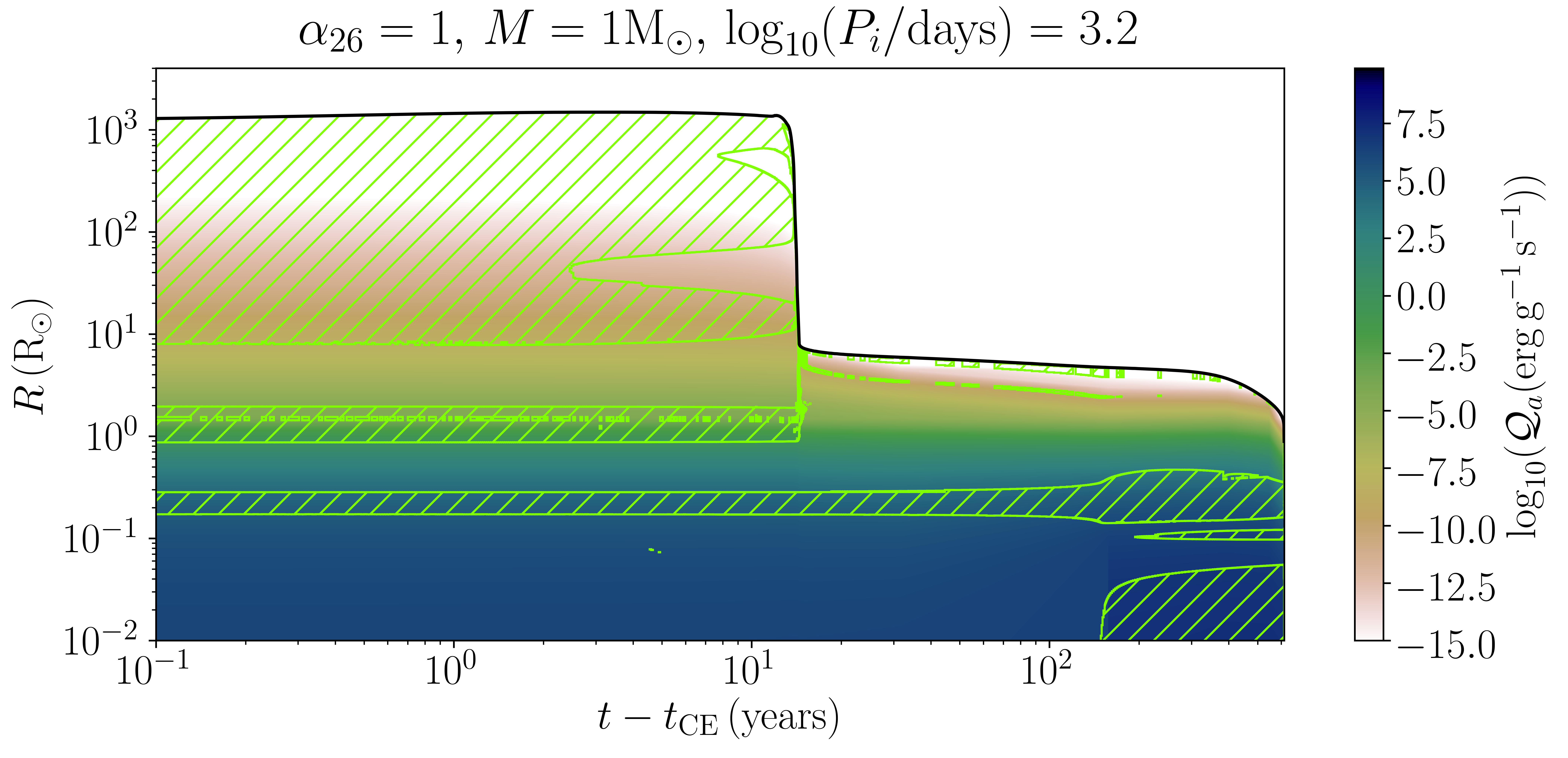}
    \caption{Axion losses during the CE phase.~The right edge corresponds to the time of merger.~Green hatching denotes convective regions.
    }
    \label{fig:ae_losses}
\end{figure}

Interestingly, we find that the axion models which do not eject their CE have enhanced nuclear burning rates compared to the SM models that do, as we demonstrate in Fig.~\ref{fig:nucburning}.~This includes strong core carbon burning during the CE phase, and is consistent with the higher temperatures in the presence of losses expected based on the homologous scaling estimates in equation \eqref{eq:homologyscaling}.~
We also find that axion emissions lead to enhanced convective regions.
This is seen in Fig.~\ref{fig:ae_losses}, and is expected from to the temperature gradient induced by the higher burning rates and the new stellar losses.~Convection has been associated with less effective CE ejection independently of axion investigations \cite{Wilson:2022elt}.

The enhanced nuclear burning in the stars which emit axions stops their contraction into the RL before the CE ejection criterion is reached.~We confirmed that this is ultimately responsible for the change in outcome by simulating an axion emitting star with an artificially small carbon burning rate.~Without a carbon burning core, the star ejected its CE.~

%%%%%%%%%%%%%%%%%%%%%%%%%
\section{Discussion and Conclusions}
\label{sec:conclusions}
%%%%%%%%%%%%%%%%%%%%%%%%%
In this work, we have studied the effects of new particle emission on the evolution of stellar binaries.~
As a test case for new particles, we have focused on light axions coupling to either electrons or photons in the stellar material.~Focusing on the case of a $30\msun$, $Z=0.02$ star with a light ($m_2\le5\msun$) BH companion, we found that new particle loss mechanisms can cause the common envelope phase to be significantly changed, delayed, or absent altogether.~This has important consequences for the BBH systems which can be observed through gravitational waves.~In particular, light particle emission can thwart the formation of BBH mergers via the common envelope pathway in two ways.~
In some systems, axion emission prevents RLOF altogether, so the objects never interact.~In others, a CE can be formed but is prevented from being ejected due to axion emission.~In the latter, we found that axion emission enhances nuclear burning rates, halting contraction and CE ejection.~This results in a merger before the star can collapse to form a BH, and primarily affects the formation of binaries in which the secondary object has a mass $ \leq 4 \msun$, within the lower mass gap.~

The parameters considered in this work are probed by other stellar objects, including the Sun (CAST) \cite{CAST:2017uph,Barth:2013sma}, which constrains $g_{10}$ and $ g_{10} \sqrt{\alpha_{26}}$ for $ m_a < 10^{-2} $ eV; horizontal branch stars \cite{Raffelt:2006cw,Ayala:2014pea,Carenza:2020zil}, which constrain $g_{10}$; and the tip of the red giant branch \cite{Viaux:2013lha,Straniero:2018fbv,Diaz:2019kim,Capozzi:2020cbu}, which constrains $\alpha_{26}$.~Until recently, the parameters we studied would have been incompatible with the stellar constraints, but recent work \cite{Dennis:2023kfe,Dennis:2023ldw} has demonstrated that the stellar bounds are significantly weakened once degeneracies and uncertainties from stellar physics are consistently accounted for in the data analyses.

This work is a preliminary study, focusing on a single stellar mass and metallicity.~A thorough exploration of the degeneracies with mass, metallicity, and other stellar parameters (including e.g.,~the mixing length) is needed to ascertain whether our conclusions are valid more generally.~Additionally, it would be interesting to explore other stellar energy loss mechanisms, as well as changes to the stellar equation of state \cite{Sakstein:2022tby}.~

Our simulations were performed using the one-dimensional code MESA, which is limited in its simulations of non-spherical systems such as binaries.~Semi-analytic prescriptions are used to calculate the effects of binary interactions, which inevitably require the introduction of free parameters describing the efficiencies of binary processes.~We have verified that our conclusions are robust to varying these parameters.~While we found small individual variations in binaries with particular initial conditions, the qualitative effects of axions were found to persist.~For example, increasing the efficiency of CE ejection increases the number of systems that eject their CE and merge within a Hubble time in the SM, but all of these systems failed to eject their CE if sufficiently strongly coupled axions are present.

From the study in this work we conclude that axion emission prevents the formation of binary black holes in the LIGO/Virgo/KAGRA lower mass gap via the common envelope BBH formation channel, at least for progenitors composed of a $30\msun$, $Z=0.02$ star and a light black hole.~To ascertain if this conclusion holds more generally, future work should simulate a larger range of masses and metallicities, include other light particle emission mechanisms, and study how light particle emission affects other lower mass gap formation channels.~We stress that it is not currently possible to draw any conclusions about the axion parameter space using compact object populations.

This work has demonstrated for the first time that light particle emission may fundamentally change stellar physics in binaries, and it would be interesting to devise novel probes of light particles using observations of binary systems.~In the era of precision gravitational wave astronomy, the effects of light particles on binary mergers cannot be ignored.

\section*{Software}
MESA version~15140, MESASDK version 20210401, Mathematica version 12.0, mkipp\footnote{\href{https://github.com/orlox/mkipp}{https://github.com/orlox/mkipp}.}.

\section*{Acknowledgements}

We thank Mitchell T.~Dennis and Sam McDermott for useful discussions.~
We are grateful to Robert Farmer, Adam Jermyn, Pablo Marchant, Mathieu Renzo, and Frank Timmes for answering our many MESA-related questions.
DC thanks her wonderful family for support while finishing this paper in the last month of her pregnancy.
DC is supported by the STFC under Grant No.~ST/T001011/1.~This material is based upon work supported by the National Science Foundation under Grant No.~2207880.
Our simulations were run on the University of Hawai\okina i's high-performance supercomputer MANA.~The technical support and advanced computing resources from University of Hawai\okina i Information Technology Services – Cyberinfrastructure, funded in part by the National Science Foundation MRI award \#1920304, are gratefully acknowledged.

\appendix

\section{Details of the MESA Simulations}
\label{appendix:MESA}

Relevant prescriptions for the MESA simulations performed as part of this work are as follows.~Convection is treated according to the Cox prescription for mixing length theory \cite{1968pss..book.....C} with mixing length parameter $\alpha_{\rm MLT}=2.0$.~Semiconvection is modeled according to  \cite{1985A&A...145..179L} with efficiency parameter $\alpha_{\rm SC}=1.0$.~We include convective overshooting of the hydrogen burning convective core using a step overshooting scheme where the size of the core is extended by $f_{\rm ov}=0.345$ pressure scale heights \cite{Brott:2011ni}.~Convective overshooting from all other regions is described using an exponential profile with exponential decay length scale $f_{\rm ov}=0.01$.~Our prescription for mass loss due to stellar winds follows that of \cite{Brott:2011ni}.~Finally, we use the MESA default nuclear burning rates (these are a mixture of the {NACRE} \cite{Angulo:1999zz} and {REACLIB} \cite{2010ApJS..189..240C} tables).~Our simulations can be reproduced using the inlists in our reproduction package \cite{djuna_croon_2022_6949679}.

\section{Stellar profiles}
\label{app:Sevol}
\begin{figure*}[ht]
    \centering
\includegraphics[width=0.4\textwidth]{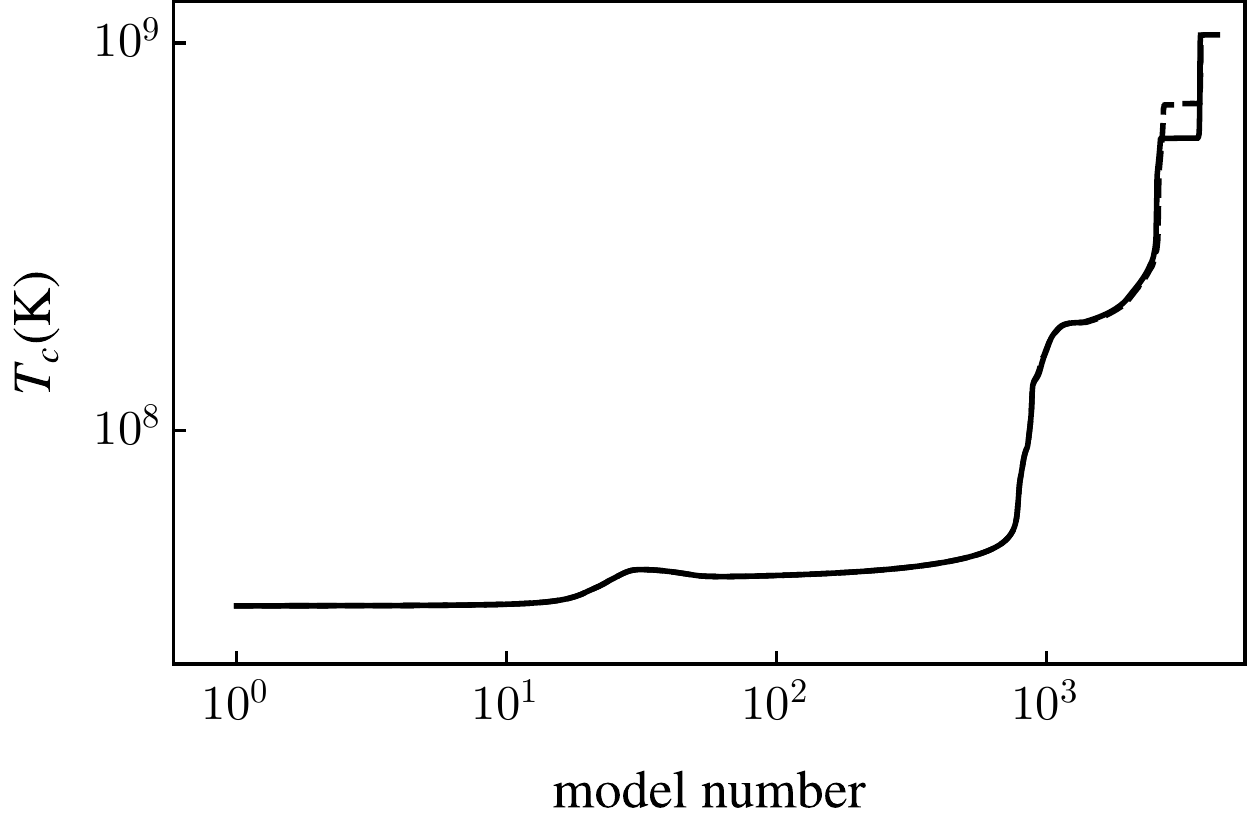}
\includegraphics[width=0.4\textwidth]{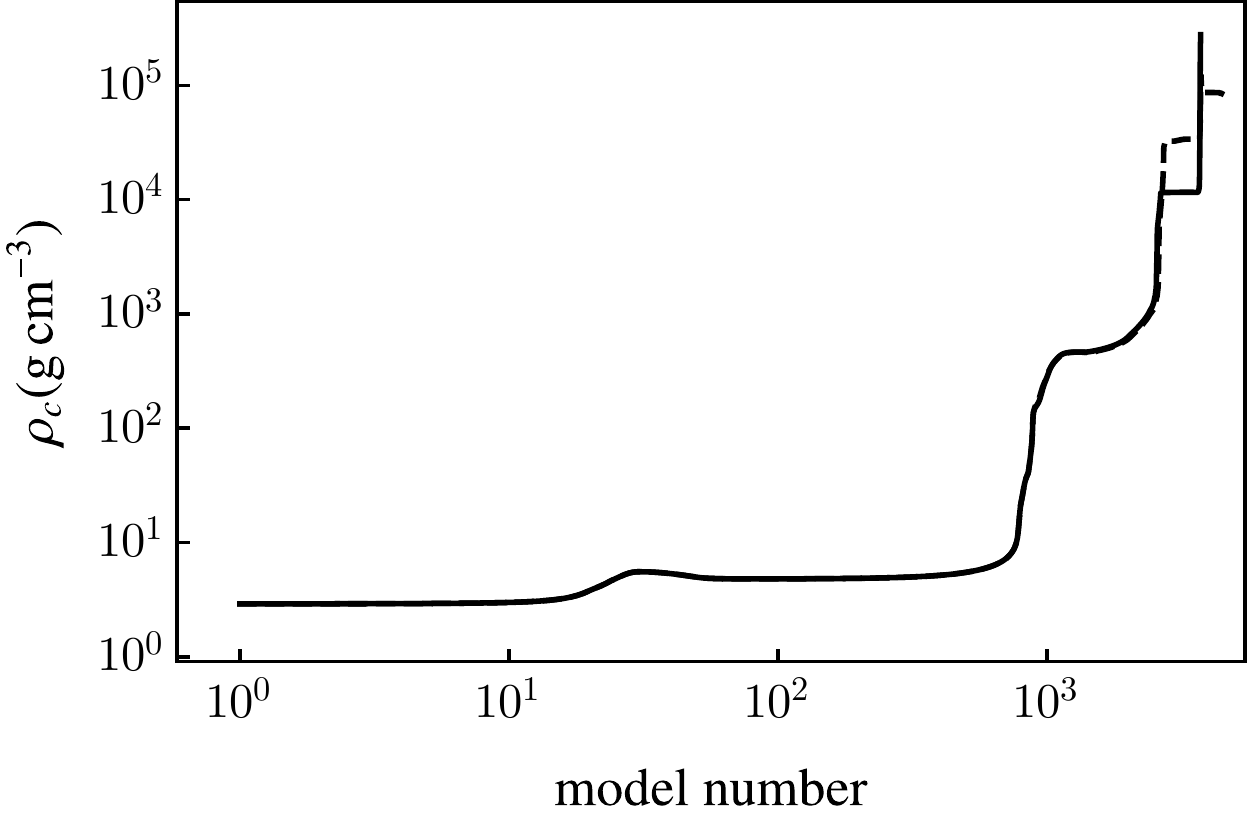}
\includegraphics[width=0.4\textwidth]{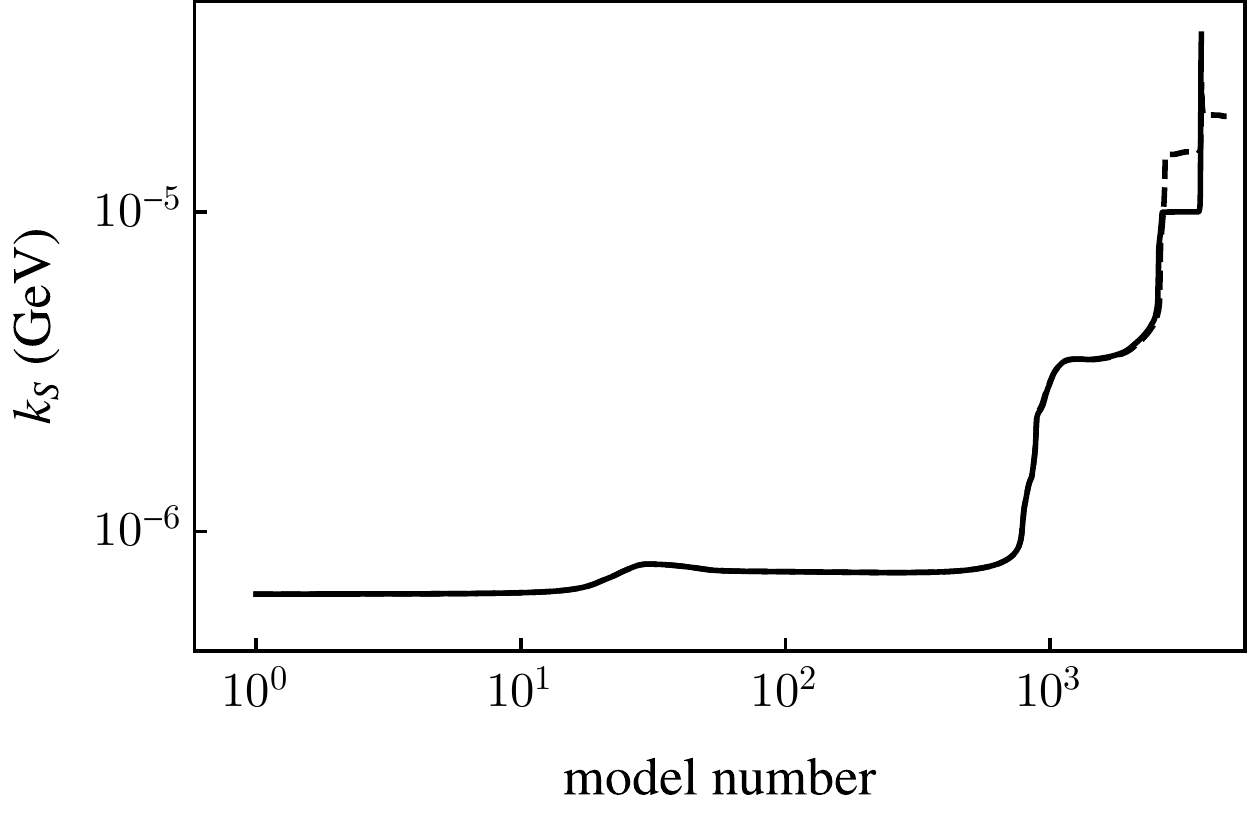}
    \caption{Core temperature (upper left), core density (upper right), and Debye screening length (lower)  as a function of model number for the stars studied in this work. The continuous lines correspond to the SM and the dashed lines correspond to the $\alpha_{26}=1 $ model.}
\label{fig:examples}
\end{figure*}

To aid the reader's intuition of the particle processes occurring in the stars studied by this work, figure \ref{fig:examples} shows example plots of the core temperature, core density, and Debye screening length as a function of model number (not linearly related to physical time).

\section{Axion Bremsstrahlung Processes}
\label{app:Brem}

We have not included axion production via bremsstrahlung processes in our simulations because the specific energy loss due to $e+ (Z,A) \to e + (Z,A) + a$ and $e + e \to e +e +a $ is only expected to become more important at higher densities than are reached by the stars we simulate in this work.~We briefly comment on bremsstrahlung processes here, both for completeness, and because our reproduction package \cite{djuna_croon_2022_6949679} includes the option for users to include them in systems where they dominate.~

\begin{spacing}{0.99}
Assuming that the electrons are nonrelativistic, the axionic bremsstrahlung rate in the non-degenerate (ND) and degenerate (D) regimes is \cite{Raffelt:1994ry}
\begin{align}
    \label{eq:Q_bremmND}
   \mathcal{Q}_{\rm ND} &= 0.58 \alpha_{26} \rho_3 T_8^{5/2} F_{b,{\rm ND}} \textrm{ ergs/g/s},
    \\ \label{eq:Q_bremmD}
    \mathcal{Q}_{\rm D}&= 10.8\, \alpha_{26}  T_8^4 F_{b,{\rm D}}\textrm{ ergs/g/s},
\end{align}
where 
\begin{equation}\scalebox{.95}{$
    F_{b,{\rm ND}} = \left(\sum_{i}\frac{X_iZ_i}{A_i}\right)\left(\sum_{i}\frac{X_iZ_i^2}{A_i}\right)+\frac{1}{\sqrt{2}}\left(\sum_{i}\frac{X_iZ_i}{A_i}\right)^2$}
\end{equation}
with $X_i$, $Z_i$, and $A_i$ the mass fraction, atomic number, and mass number of species $i$ respectively.~The sum runs over all ion species.~To second order in the velocity at the Fermi surface $\beta_F = p_F/E_F$,
\begin{equation}\scalebox{.9}{$
F_{b,{\rm D}} = \frac{2}{3} \log\left(\frac{2+\kappa^2}{\kappa^2} \right) + \left[ \left(\kappa^2 +\frac25 \right) \log\left( \frac{2+\kappa^2}{\kappa^2} \right) - 2\right] \frac{\beta_F^2}3,$}
\end{equation}
where the Debye angle is $\kappa^2 = k_S^2/(2p_F^2)$ and the Debye momentum is given by \eqref{eq:debyeMomentum}.~The axion loss rates due to Bremsstrahlung processes are implemented into MESA using the interpolating formula $( \mathcal Q_{b,{\rm ND}}^{-1} + \mathcal Q_{b,{\rm D}}^{-1})^{-1}$ \cite{Raffelt:1994ry}.
\end{spacing}

\bibliography{refs}

\end{document}

%% file: universalnewcommands.tex
\newcommand{\beq}{\begin{equation}}
\newcommand{\eeq}{\end{equation}}
\newcommand{\bea}{\begin{eqnarray}}
\newcommand{\eea}{\end{eqnarray}}

     \newcommand{\cL}{{\cal L}}

\newcommand{\refjnl}[1]{{\rm#1}}

\def\apjs{\refjnl{ApJS}}               % Astrophysical Journal, Supplement
\def\aap{\refjnl{A\&A}}                % Astronomy and Astrophysics
\def\mnras{\refjnl{MNRAS}}             % Monthly Notices of the RAS

%% file: main_v2.bbl
 \newcommand{\noop}[1]{}
\begin{thebibliography}{63}%
\makeatletter
\providecommand \@ifxundefined [1]{%
 \@ifx{#1\undefined}
}%
\providecommand \@ifnum [1]{%
 \ifnum #1\expandafter \@firstoftwo
 \else \expandafter \@secondoftwo
 \fi
}%
\providecommand \@ifx [1]{%
 \ifx #1\expandafter \@firstoftwo
 \else \expandafter \@secondoftwo
 \fi
}%
\providecommand \natexlab [1]{#1}%
\providecommand \enquote  [1]{``#1''}%
\providecommand \bibnamefont  [1]{#1}%
\providecommand \bibfnamefont [1]{#1}%
\providecommand \citenamefont [1]{#1}%
\providecommand \href@noop [0]{\@secondoftwo}%
\providecommand \href [0]{\begingroup \@sanitize@url \@href}%
\providecommand \@href[1]{\@@startlink{#1}\@@href}%
\providecommand \@@href[1]{\endgroup#1\@@endlink}%
\providecommand \@sanitize@url [0]{\catcode `\\12\catcode `\$12\catcode
  `\&12\catcode `\#12\catcode `\^12\catcode `\_12\catcode `\%12\relax}%
\providecommand \@@startlink[1]{}%
\providecommand \@@endlink[0]{}%
\providecommand \url  [0]{\begingroup\@sanitize@url \@url }%
\providecommand \@url [1]{\endgroup\@href {#1}{\urlprefix }}%
\providecommand \urlprefix  [0]{URL }%
\providecommand \Eprint [0]{\href }%
\providecommand \doibase [0]{http://dx.doi.org/}%
\providecommand \selectlanguage [0]{\@gobble}%
\providecommand \bibinfo  [0]{\@secondoftwo}%
\providecommand \bibfield  [0]{\@secondoftwo}%
\providecommand \translation [1]{[#1]}%
\providecommand \BibitemOpen [0]{}%
\providecommand \bibitemStop [0]{}%
\providecommand \bibitemNoStop [0]{.\EOS\space}%
\providecommand \EOS [0]{\spacefactor3000\relax}%
\providecommand \BibitemShut  [1]{\csname bibitem#1\endcsname}%
\let\auto@bib@innerbib\@empty
%</preamble>
\bibitem [{\citenamefont {Abbott}\ \emph
  {et~al.}(2021{\natexlab{a}})\citenamefont {Abbott} \emph
  {et~al.}}]{LIGOScientific:2021djp}%
  \BibitemOpen
  \bibfield  {author} {\bibinfo {author} {\bibfnamefont {R.}~\bibnamefont
  {Abbott}} \emph {et~al.} (\bibinfo {collaboration} {LIGO Scientific, VIRGO,
  KAGRA}),\ }\href@noop {} {\  (\bibinfo {year} {2021}{\natexlab{a}})},\
  \Eprint {http://arxiv.org/abs/2111.03606} {arXiv:2111.03606 [gr-qc]}
  \BibitemShut {NoStop}%
\bibitem [{\citenamefont {Abbott}\ \emph
  {et~al.}(2021{\natexlab{b}})\citenamefont {Abbott} \emph
  {et~al.}}]{LIGOScientific:2021psn}%
  \BibitemOpen
  \bibfield  {author} {\bibinfo {author} {\bibfnamefont {R.}~\bibnamefont
  {Abbott}} \emph {et~al.} (\bibinfo {collaboration} {LIGO Scientific, VIRGO,
  KAGRA}),\ }\href@noop {} {\  (\bibinfo {year} {2021}{\natexlab{b}})},\
  \Eprint {http://arxiv.org/abs/2111.03634} {arXiv:2111.03634 [astro-ph.HE]}
  \BibitemShut {NoStop}%
\bibitem [{\citenamefont {Straight}\ \emph {et~al.}(2020)\citenamefont
  {Straight}, \citenamefont {Sakstein},\ and\ \citenamefont
  {Baxter}}]{Straight:2020zke}%
  \BibitemOpen
  \bibfield  {author} {\bibinfo {author} {\bibfnamefont {M.~C.}\ \bibnamefont
  {Straight}}, \bibinfo {author} {\bibfnamefont {J.}~\bibnamefont {Sakstein}},
  \ and\ \bibinfo {author} {\bibfnamefont {E.~J.}\ \bibnamefont {Baxter}},\
  }\href {\doibase 10.1103/PhysRevD.102.124018} {\bibfield  {journal} {\bibinfo
   {journal} {Phys. Rev. D}\ }\textbf {\bibinfo {volume} {102}},\ \bibinfo
  {pages} {124018} (\bibinfo {year} {2020})},\ \Eprint
  {http://arxiv.org/abs/2009.10716} {arXiv:2009.10716 [gr-qc]} \BibitemShut
  {NoStop}%
\bibitem [{\citenamefont {Farmer}\ \emph {et~al.}(2020)\citenamefont {Farmer},
  \citenamefont {Renzo}, \citenamefont {de~Mink}, \citenamefont {Fishbach},\
  and\ \citenamefont {Justham}}]{Farmer:2020xne}%
  \BibitemOpen
  \bibfield  {author} {\bibinfo {author} {\bibfnamefont {R.}~\bibnamefont
  {Farmer}}, \bibinfo {author} {\bibfnamefont {M.}~\bibnamefont {Renzo}},
  \bibinfo {author} {\bibfnamefont {S.}~\bibnamefont {de~Mink}}, \bibinfo
  {author} {\bibfnamefont {M.}~\bibnamefont {Fishbach}}, \ and\ \bibinfo
  {author} {\bibfnamefont {S.}~\bibnamefont {Justham}},\ }\href@noop {} {\
  (\bibinfo {year} {2020})},\ \Eprint {http://arxiv.org/abs/2006.06678}
  {arXiv:2006.06678 [astro-ph.HE]} \BibitemShut {NoStop}%
\bibitem [{\citenamefont {Croon}\ \emph
  {et~al.}(2020{\natexlab{a}})\citenamefont {Croon}, \citenamefont
  {McDermott},\ and\ \citenamefont {Sakstein}}]{Croon:2020oga}%
  \BibitemOpen
  \bibfield  {author} {\bibinfo {author} {\bibfnamefont {D.}~\bibnamefont
  {Croon}}, \bibinfo {author} {\bibfnamefont {S.~D.}\ \bibnamefont
  {McDermott}}, \ and\ \bibinfo {author} {\bibfnamefont {J.}~\bibnamefont
  {Sakstein}},\ }\href@noop {} {\  (\bibinfo {year} {2020}{\natexlab{a}})},\
  \Eprint {http://arxiv.org/abs/2007.07889} {arXiv:2007.07889 [gr-qc]}
  \BibitemShut {NoStop}%
\bibitem [{\citenamefont {Sakstein}\ \emph {et~al.}(2020)\citenamefont
  {Sakstein}, \citenamefont {Croon}, \citenamefont {McDermott}, \citenamefont
  {Straight},\ and\ \citenamefont {Baxter}}]{Sakstein:2020axg}%
  \BibitemOpen
  \bibfield  {author} {\bibinfo {author} {\bibfnamefont {J.}~\bibnamefont
  {Sakstein}}, \bibinfo {author} {\bibfnamefont {D.}~\bibnamefont {Croon}},
  \bibinfo {author} {\bibfnamefont {S.~D.}\ \bibnamefont {McDermott}}, \bibinfo
  {author} {\bibfnamefont {M.~C.}\ \bibnamefont {Straight}}, \ and\ \bibinfo
  {author} {\bibfnamefont {E.~J.}\ \bibnamefont {Baxter}},\ }\href {\doibase
  10.1103/PhysRevLett.125.261105} {\bibfield  {journal} {\bibinfo  {journal}
  {Phys. Rev. Lett.}\ }\textbf {\bibinfo {volume} {125}},\ \bibinfo {pages}
  {261105} (\bibinfo {year} {2020})},\ \Eprint
  {http://arxiv.org/abs/2009.01213} {arXiv:2009.01213 [gr-qc]} \BibitemShut
  {NoStop}%
\bibitem [{\citenamefont {Baxter}\ \emph {et~al.}(2021)\citenamefont {Baxter},
  \citenamefont {Croon}, \citenamefont {McDermott},\ and\ \citenamefont
  {Sakstein}}]{Baxter:2021swn}%
  \BibitemOpen
  \bibfield  {author} {\bibinfo {author} {\bibfnamefont {E.~J.}\ \bibnamefont
  {Baxter}}, \bibinfo {author} {\bibfnamefont {D.}~\bibnamefont {Croon}},
  \bibinfo {author} {\bibfnamefont {S.~D.}\ \bibnamefont {McDermott}}, \ and\
  \bibinfo {author} {\bibfnamefont {J.}~\bibnamefont {Sakstein}},\ }\href
  {\doibase 10.3847/2041-8213/ac11fc} {\bibfield  {journal} {\bibinfo
  {journal} {Astrophys. J. Lett.}\ }\textbf {\bibinfo {volume} {916}},\
  \bibinfo {pages} {L16} (\bibinfo {year} {2021})},\ \Eprint
  {http://arxiv.org/abs/2104.02685} {arXiv:2104.02685 [astro-ph.CO]}
  \BibitemShut {NoStop}%
\bibitem [{\citenamefont {Paczynski}\ \emph {et~al.}(1976)\citenamefont
  {Paczynski}, \citenamefont {Eggleton}, \citenamefont {Mitton},\ and\
  \citenamefont {Whelan}}]{paczynski1976structure}%
  \BibitemOpen
  \bibfield  {author} {\bibinfo {author} {\bibfnamefont {B.}~\bibnamefont
  {Paczynski}}, \bibinfo {author} {\bibfnamefont {P.}~\bibnamefont {Eggleton}},
  \bibinfo {author} {\bibfnamefont {S.}~\bibnamefont {Mitton}}, \ and\ \bibinfo
  {author} {\bibfnamefont {J.}~\bibnamefont {Whelan}},\ }in\ \href@noop {}
  {\emph {\bibinfo {booktitle} {IAU Symp}}},\ Vol.~\bibinfo {volume} {73}\
  (\bibinfo {year} {1976})\ p.~\bibinfo {pages} {75}\BibitemShut {NoStop}%
\bibitem [{\citenamefont {Van~den Heuvel}\ and\ \citenamefont
  {Eggleton}(1976)}]{van1976structure}%
  \BibitemOpen
  \bibfield  {author} {\bibinfo {author} {\bibfnamefont {E.}~\bibnamefont
  {Van~den Heuvel}}\ and\ \bibinfo {author} {\bibfnamefont {P.}~\bibnamefont
  {Eggleton}},\ }in\ \href@noop {} {\emph {\bibinfo {booktitle} {IAU Symp}}},\
  Vol.~\bibinfo {volume} {73}\ (\bibinfo {year} {1976})\ pp.\ \bibinfo {pages}
  {35--61}\BibitemShut {NoStop}%
\bibitem [{\citenamefont {Tutukov}\ and\ \citenamefont
  {Yungelson}(1993)}]{tutukov1993merger}%
  \BibitemOpen
  \bibfield  {author} {\bibinfo {author} {\bibfnamefont {A.}~\bibnamefont
  {Tutukov}}\ and\ \bibinfo {author} {\bibfnamefont {L.}~\bibnamefont
  {Yungelson}},\ }\href@noop {} {\bibfield  {journal} {\bibinfo  {journal}
  {Monthly Notices of the Royal Astronomical Society}\ }\textbf {\bibinfo
  {volume} {260}},\ \bibinfo {pages} {675} (\bibinfo {year}
  {1993})}\BibitemShut {NoStop}%
\bibitem [{\citenamefont {Hurley}\ \emph {et~al.}(2002)\citenamefont {Hurley},
  \citenamefont {Tout},\ and\ \citenamefont {Pols}}]{hurley2002evolution}%
  \BibitemOpen
  \bibfield  {author} {\bibinfo {author} {\bibfnamefont {J.~R.}\ \bibnamefont
  {Hurley}}, \bibinfo {author} {\bibfnamefont {C.~A.}\ \bibnamefont {Tout}}, \
  and\ \bibinfo {author} {\bibfnamefont {O.~R.}\ \bibnamefont {Pols}},\
  }\href@noop {} {\bibfield  {journal} {\bibinfo  {journal} {Monthly Notices of
  the Royal Astronomical Society}\ }\textbf {\bibinfo {volume} {329}},\
  \bibinfo {pages} {897} (\bibinfo {year} {2002})}\BibitemShut {NoStop}%
\bibitem [{\citenamefont {Ivanova}(2011)}]{Ivanova:2011tc}%
  \BibitemOpen
  \bibfield  {author} {\bibinfo {author} {\bibfnamefont {N.}~\bibnamefont
  {Ivanova}},\ }\href {\doibase 10.1088/0004-637X/730/2/76} {\bibfield
  {journal} {\bibinfo  {journal} {Astrophys. J.}\ }\textbf {\bibinfo {volume}
  {730}},\ \bibinfo {pages} {76} (\bibinfo {year} {2011})},\ \Eprint
  {http://arxiv.org/abs/1101.2863} {arXiv:1101.2863 [astro-ph.SR]} \BibitemShut
  {NoStop}%
\bibitem [{\citenamefont {Ivanova}\ \emph {et~al.}(2013)\citenamefont
  {Ivanova}, \citenamefont {Justham}, \citenamefont {Chen}, \citenamefont
  {De~Marco}, \citenamefont {Fryer}, \citenamefont {Gaburov}, \citenamefont
  {Ge}, \citenamefont {Glebbeek}, \citenamefont {Han}, \citenamefont {Li} \emph
  {et~al.}}]{ivanova2013common}%
  \BibitemOpen
  \bibfield  {author} {\bibinfo {author} {\bibfnamefont {N.}~\bibnamefont
  {Ivanova}}, \bibinfo {author} {\bibfnamefont {S.}~\bibnamefont {Justham}},
  \bibinfo {author} {\bibfnamefont {X.}~\bibnamefont {Chen}}, \bibinfo {author}
  {\bibfnamefont {O.}~\bibnamefont {De~Marco}}, \bibinfo {author}
  {\bibfnamefont {C.}~\bibnamefont {Fryer}}, \bibinfo {author} {\bibfnamefont
  {E.}~\bibnamefont {Gaburov}}, \bibinfo {author} {\bibfnamefont
  {H.}~\bibnamefont {Ge}}, \bibinfo {author} {\bibfnamefont {E.}~\bibnamefont
  {Glebbeek}}, \bibinfo {author} {\bibfnamefont {Z.}~\bibnamefont {Han}},
  \bibinfo {author} {\bibfnamefont {X.-D.}\ \bibnamefont {Li}},  \emph
  {et~al.},\ }\href@noop {} {\bibfield  {journal} {\bibinfo  {journal} {The
  Astronomy and Astrophysics Review}\ }\textbf {\bibinfo {volume} {21}},\
  \bibinfo {pages} {1} (\bibinfo {year} {2013})}\BibitemShut {NoStop}%
\bibitem [{\citenamefont {Mapelli}(2020)}]{mapelli2020binary}%
  \BibitemOpen
  \bibfield  {author} {\bibinfo {author} {\bibfnamefont {M.}~\bibnamefont
  {Mapelli}},\ }\href@noop {} {\bibfield  {journal} {\bibinfo  {journal}
  {Frontiers in Astronomy and Space Sciences}\ }\textbf {\bibinfo {volume}
  {7}},\ \bibinfo {pages} {38} (\bibinfo {year} {2020})}\BibitemShut {NoStop}%
\bibitem [{\citenamefont {Kreidberg}\ \emph {et~al.}(2012)\citenamefont
  {Kreidberg}, \citenamefont {Bailyn}, \citenamefont {Farr},\ and\
  \citenamefont {Kalogera}}]{Kreidberg:2012ud}%
  \BibitemOpen
  \bibfield  {author} {\bibinfo {author} {\bibfnamefont {L.}~\bibnamefont
  {Kreidberg}}, \bibinfo {author} {\bibfnamefont {C.~D.}\ \bibnamefont
  {Bailyn}}, \bibinfo {author} {\bibfnamefont {W.~M.}\ \bibnamefont {Farr}}, \
  and\ \bibinfo {author} {\bibfnamefont {V.}~\bibnamefont {Kalogera}},\ }\href
  {\doibase 10.1088/0004-637X/757/1/36} {\bibfield  {journal} {\bibinfo
  {journal} {Astrophys. J.}\ }\textbf {\bibinfo {volume} {757}},\ \bibinfo
  {pages} {36} (\bibinfo {year} {2012})},\ \Eprint
  {http://arxiv.org/abs/1205.1805} {arXiv:1205.1805 [astro-ph.HE]} \BibitemShut
  {NoStop}%
\bibitem [{\citenamefont {Abbott}\ \emph {et~al.}(2020)\citenamefont {Abbott}
  \emph {et~al.}}]{LIGOScientific:2020zkf}%
  \BibitemOpen
  \bibfield  {author} {\bibinfo {author} {\bibfnamefont {R.}~\bibnamefont
  {Abbott}} \emph {et~al.} (\bibinfo {collaboration} {LIGO Scientific,
  Virgo}),\ }\href {\doibase 10.3847/2041-8213/ab960f} {\bibfield  {journal}
  {\bibinfo  {journal} {Astrophys. J. Lett.}\ }\textbf {\bibinfo {volume}
  {896}},\ \bibinfo {pages} {L44} (\bibinfo {year} {2020})},\ \Eprint
  {http://arxiv.org/abs/2006.12611} {arXiv:2006.12611 [astro-ph.HE]}
  \BibitemShut {NoStop}%
\bibitem [{\citenamefont {Zevin}\ \emph {et~al.}(2021)\citenamefont {Zevin},
  \citenamefont {Bavera}, \citenamefont {Berry}, \citenamefont {Kalogera},
  \citenamefont {Fragos}, \citenamefont {Marchant}, \citenamefont {Rodriguez},
  \citenamefont {Antonini}, \citenamefont {Holz},\ and\ \citenamefont
  {Pankow}}]{Zevin:2020gbd}%
  \BibitemOpen
  \bibfield  {author} {\bibinfo {author} {\bibfnamefont {M.}~\bibnamefont
  {Zevin}}, \bibinfo {author} {\bibfnamefont {S.~S.}\ \bibnamefont {Bavera}},
  \bibinfo {author} {\bibfnamefont {C.~P.~L.}\ \bibnamefont {Berry}}, \bibinfo
  {author} {\bibfnamefont {V.}~\bibnamefont {Kalogera}}, \bibinfo {author}
  {\bibfnamefont {T.}~\bibnamefont {Fragos}}, \bibinfo {author} {\bibfnamefont
  {P.}~\bibnamefont {Marchant}}, \bibinfo {author} {\bibfnamefont {C.~L.}\
  \bibnamefont {Rodriguez}}, \bibinfo {author} {\bibfnamefont {F.}~\bibnamefont
  {Antonini}}, \bibinfo {author} {\bibfnamefont {D.~E.}\ \bibnamefont {Holz}},
  \ and\ \bibinfo {author} {\bibfnamefont {C.}~\bibnamefont {Pankow}},\ }\href
  {\doibase 10.3847/1538-4357/abe40e} {\bibfield  {journal} {\bibinfo
  {journal} {Astrophys. J.}\ }\textbf {\bibinfo {volume} {910}},\ \bibinfo
  {pages} {152} (\bibinfo {year} {2021})},\ \Eprint
  {http://arxiv.org/abs/2011.10057} {arXiv:2011.10057 [astro-ph.HE]}
  \BibitemShut {NoStop}%
\bibitem [{\citenamefont {van~den Heuvel}\ \emph {et~al.}(2017)\citenamefont
  {van~den Heuvel}, \citenamefont {Portegies~Zwart},\ and\ \citenamefont
  {de~Mink}}]{van2017forming}%
  \BibitemOpen
  \bibfield  {author} {\bibinfo {author} {\bibfnamefont {E.~P.}\ \bibnamefont
  {van~den Heuvel}}, \bibinfo {author} {\bibfnamefont {S.}~\bibnamefont
  {Portegies~Zwart}}, \ and\ \bibinfo {author} {\bibfnamefont {S.~E.}\
  \bibnamefont {de~Mink}},\ }\href@noop {} {\bibfield  {journal} {\bibinfo
  {journal} {Monthly Notices of The Royal Astronomical Society}\ }\textbf
  {\bibinfo {volume} {471}},\ \bibinfo {pages} {4256} (\bibinfo {year}
  {2017})}\BibitemShut {NoStop}%
\bibitem [{\citenamefont {Neijssel}\ \emph {et~al.}(2019)\citenamefont
  {Neijssel}, \citenamefont {Vigna-G{\'o}mez}, \citenamefont {Stevenson},
  \citenamefont {Barrett}, \citenamefont {Gaebel}, \citenamefont
  {Broekgaarden}, \citenamefont {de~Mink}, \citenamefont {Sz{\'e}csi},
  \citenamefont {Vinciguerra},\ and\ \citenamefont
  {Mandel}}]{neijssel2019effect}%
  \BibitemOpen
  \bibfield  {author} {\bibinfo {author} {\bibfnamefont {C.~J.}\ \bibnamefont
  {Neijssel}}, \bibinfo {author} {\bibfnamefont {A.}~\bibnamefont
  {Vigna-G{\'o}mez}}, \bibinfo {author} {\bibfnamefont {S.}~\bibnamefont
  {Stevenson}}, \bibinfo {author} {\bibfnamefont {J.~W.}\ \bibnamefont
  {Barrett}}, \bibinfo {author} {\bibfnamefont {S.~M.}\ \bibnamefont {Gaebel}},
  \bibinfo {author} {\bibfnamefont {F.~S.}\ \bibnamefont {Broekgaarden}},
  \bibinfo {author} {\bibfnamefont {S.~E.}\ \bibnamefont {de~Mink}}, \bibinfo
  {author} {\bibfnamefont {D.}~\bibnamefont {Sz{\'e}csi}}, \bibinfo {author}
  {\bibfnamefont {S.}~\bibnamefont {Vinciguerra}}, \ and\ \bibinfo {author}
  {\bibfnamefont {I.}~\bibnamefont {Mandel}},\ }\href@noop {} {\bibfield
  {journal} {\bibinfo  {journal} {Monthly Notices of the Royal Astronomical
  Society}\ }\textbf {\bibinfo {volume} {490}},\ \bibinfo {pages} {3740}
  (\bibinfo {year} {2019})}\BibitemShut {NoStop}%
\bibitem [{\citenamefont {Marchant}\ \emph {et~al.}(2016)\citenamefont
  {Marchant}, \citenamefont {Langer}, \citenamefont {Podsiadlowski},
  \citenamefont {Tauris},\ and\ \citenamefont {Moriya}}]{Marchant:2016wow}%
  \BibitemOpen
  \bibfield  {author} {\bibinfo {author} {\bibfnamefont {P.}~\bibnamefont
  {Marchant}}, \bibinfo {author} {\bibfnamefont {N.}~\bibnamefont {Langer}},
  \bibinfo {author} {\bibfnamefont {P.}~\bibnamefont {Podsiadlowski}}, \bibinfo
  {author} {\bibfnamefont {T.~M.}\ \bibnamefont {Tauris}}, \ and\ \bibinfo
  {author} {\bibfnamefont {T.~J.}\ \bibnamefont {Moriya}},\ }\href {\doibase
  10.1051/0004-6361/201628133} {\bibfield  {journal} {\bibinfo  {journal}
  {Astron. Astrophys.}\ }\textbf {\bibinfo {volume} {588}},\ \bibinfo {pages}
  {A50} (\bibinfo {year} {2016})},\ \Eprint {http://arxiv.org/abs/1601.03718}
  {arXiv:1601.03718 [astro-ph.SR]} \BibitemShut {NoStop}%
\bibitem [{\citenamefont {{Maeder}}(1987)}]{1987A&A...178..159M}%
  \BibitemOpen
  \bibfield  {author} {\bibinfo {author} {\bibfnamefont {A.}~\bibnamefont
  {{Maeder}}},\ }\href@noop {} {\bibfield  {journal} {\bibinfo  {journal}
  {\aap}\ }\textbf {\bibinfo {volume} {178}},\ \bibinfo {pages} {159} (\bibinfo
  {year} {1987})}\BibitemShut {NoStop}%
\bibitem [{\citenamefont {Mandel}\ and\ \citenamefont
  {De~Mink}(2016)}]{mandel2016merging}%
  \BibitemOpen
  \bibfield  {author} {\bibinfo {author} {\bibfnamefont {I.}~\bibnamefont
  {Mandel}}\ and\ \bibinfo {author} {\bibfnamefont {S.~E.}\ \bibnamefont
  {De~Mink}},\ }\href@noop {} {\bibfield  {journal} {\bibinfo  {journal}
  {Monthly Notices of the Royal Astronomical Society}\ }\textbf {\bibinfo
  {volume} {458}},\ \bibinfo {pages} {2634} (\bibinfo {year}
  {2016})}\BibitemShut {NoStop}%
\bibitem [{\citenamefont {Thompson}(2011)}]{thompson2011accelerating}%
  \BibitemOpen
  \bibfield  {author} {\bibinfo {author} {\bibfnamefont {T.~A.}\ \bibnamefont
  {Thompson}},\ }\href@noop {} {\bibfield  {journal} {\bibinfo  {journal} {The
  Astrophysical Journal}\ }\textbf {\bibinfo {volume} {741}},\ \bibinfo {pages}
  {82} (\bibinfo {year} {2011})}\BibitemShut {NoStop}%
\bibitem [{\citenamefont {Antonini}\ \emph {et~al.}(2017)\citenamefont
  {Antonini}, \citenamefont {Toonen},\ and\ \citenamefont
  {Hamers}}]{antonini2017binary}%
  \BibitemOpen
  \bibfield  {author} {\bibinfo {author} {\bibfnamefont {F.}~\bibnamefont
  {Antonini}}, \bibinfo {author} {\bibfnamefont {S.}~\bibnamefont {Toonen}}, \
  and\ \bibinfo {author} {\bibfnamefont {A.~S.}\ \bibnamefont {Hamers}},\
  }\href@noop {} {\bibfield  {journal} {\bibinfo  {journal} {The Astrophysical
  Journal}\ }\textbf {\bibinfo {volume} {841}},\ \bibinfo {pages} {77}
  (\bibinfo {year} {2017})}\BibitemShut {NoStop}%
\bibitem [{\citenamefont {Vigna-G\'omez}\ \emph {et~al.}(2021)\citenamefont
  {Vigna-G\'omez}, \citenamefont {Toonen}, \citenamefont {Ramirez-Ruiz},
  \citenamefont {Leigh}, \citenamefont {Riley},\ and\ \citenamefont
  {Haster}}]{Vigna-Gomez:2020fvw}%
  \BibitemOpen
  \bibfield  {author} {\bibinfo {author} {\bibfnamefont {A.}~\bibnamefont
  {Vigna-G\'omez}}, \bibinfo {author} {\bibfnamefont {S.}~\bibnamefont
  {Toonen}}, \bibinfo {author} {\bibfnamefont {E.}~\bibnamefont
  {Ramirez-Ruiz}}, \bibinfo {author} {\bibfnamefont {N.~W.~C.}\ \bibnamefont
  {Leigh}}, \bibinfo {author} {\bibfnamefont {J.}~\bibnamefont {Riley}}, \ and\
  \bibinfo {author} {\bibfnamefont {C.-J.}\ \bibnamefont {Haster}},\ }\href
  {\doibase 10.3847/2041-8213/abd5b7} {\bibfield  {journal} {\bibinfo
  {journal} {Astrophys. J. Lett.}\ }\textbf {\bibinfo {volume} {907}},\
  \bibinfo {pages} {L19} (\bibinfo {year} {2021})},\ \Eprint
  {http://arxiv.org/abs/2010.13669} {arXiv:2010.13669 [astro-ph.HE]}
  \BibitemShut {NoStop}%
\bibitem [{\citenamefont {Paxton}\ \emph {et~al.}(2011)\citenamefont {Paxton},
  \citenamefont {Bildsten}, \citenamefont {Dotter}, \citenamefont {Herwig},
  \citenamefont {Lesaffre},\ and\ \citenamefont {Timmes}}]{Paxton:2010ji}%
  \BibitemOpen
  \bibfield  {author} {\bibinfo {author} {\bibfnamefont {B.}~\bibnamefont
  {Paxton}}, \bibinfo {author} {\bibfnamefont {L.}~\bibnamefont {Bildsten}},
  \bibinfo {author} {\bibfnamefont {A.}~\bibnamefont {Dotter}}, \bibinfo
  {author} {\bibfnamefont {F.}~\bibnamefont {Herwig}}, \bibinfo {author}
  {\bibfnamefont {P.}~\bibnamefont {Lesaffre}}, \ and\ \bibinfo {author}
  {\bibfnamefont {F.}~\bibnamefont {Timmes}},\ }\href {\doibase
  10.1088/0067-0049/192/1/3} {\bibfield  {journal} {\bibinfo  {journal}
  {Astrophys. J. Suppl.}\ }\textbf {\bibinfo {volume} {192}},\ \bibinfo {pages}
  {3} (\bibinfo {year} {2011})},\ \Eprint {http://arxiv.org/abs/1009.1622}
  {arXiv:1009.1622 [astro-ph.SR]} \BibitemShut {NoStop}%
\bibitem [{\citenamefont {Paxton}\ \emph {et~al.}(2013)\citenamefont {Paxton}
  \emph {et~al.}}]{Paxton:2013pj}%
  \BibitemOpen
  \bibfield  {author} {\bibinfo {author} {\bibfnamefont {B.}~\bibnamefont
  {Paxton}} \emph {et~al.},\ }\href {\doibase 10.1088/0067-0049/208/1/4}
  {\bibfield  {journal} {\bibinfo  {journal} {Astrophys. J. Suppl.}\ }\textbf
  {\bibinfo {volume} {208}},\ \bibinfo {pages} {4} (\bibinfo {year} {2013})},\
  \Eprint {http://arxiv.org/abs/1301.0319} {arXiv:1301.0319 [astro-ph.SR]}
  \BibitemShut {NoStop}%
\bibitem [{\citenamefont {Paxton}\ \emph {et~al.}(2015)\citenamefont {Paxton}
  \emph {et~al.}}]{Paxton:2015jva}%
  \BibitemOpen
  \bibfield  {author} {\bibinfo {author} {\bibfnamefont {B.}~\bibnamefont
  {Paxton}} \emph {et~al.},\ }\href {\doibase 10.1088/0067-0049/220/1/15}
  {\bibfield  {journal} {\bibinfo  {journal} {Astrophys. J. Suppl.}\ }\textbf
  {\bibinfo {volume} {220}},\ \bibinfo {pages} {15} (\bibinfo {year} {2015})},\
  \Eprint {http://arxiv.org/abs/1506.03146} {arXiv:1506.03146 [astro-ph.SR]}
  \BibitemShut {NoStop}%
\bibitem [{\citenamefont {Paxton}\ \emph {et~al.}(2018)\citenamefont {Paxton}
  \emph {et~al.}}]{Paxton:2017eie}%
  \BibitemOpen
  \bibfield  {author} {\bibinfo {author} {\bibfnamefont {B.}~\bibnamefont
  {Paxton}} \emph {et~al.},\ }\href {\doibase 10.3847/1538-4365/aaa5a8}
  {\bibfield  {journal} {\bibinfo  {journal} {Astrophys. J. Suppl.}\ }\textbf
  {\bibinfo {volume} {234}},\ \bibinfo {pages} {34} (\bibinfo {year} {2018})},\
  \Eprint {http://arxiv.org/abs/1710.08424} {arXiv:1710.08424 [astro-ph.SR]}
  \BibitemShut {NoStop}%
\bibitem [{\citenamefont {Marchant}\ \emph {et~al.}(2021)\citenamefont
  {Marchant}, \citenamefont {Pappas}, \citenamefont {Gallegos-Garcia},
  \citenamefont {Berry}, \citenamefont {Taam}, \citenamefont {Kalogera},\ and\
  \citenamefont {Podsiadlowski}}]{Marchant:2021hiv}%
  \BibitemOpen
  \bibfield  {author} {\bibinfo {author} {\bibfnamefont {P.}~\bibnamefont
  {Marchant}}, \bibinfo {author} {\bibfnamefont {K.~M.~W.}\ \bibnamefont
  {Pappas}}, \bibinfo {author} {\bibfnamefont {M.}~\bibnamefont
  {Gallegos-Garcia}}, \bibinfo {author} {\bibfnamefont {C.~P.~L.}\ \bibnamefont
  {Berry}}, \bibinfo {author} {\bibfnamefont {R.~E.}\ \bibnamefont {Taam}},
  \bibinfo {author} {\bibfnamefont {V.}~\bibnamefont {Kalogera}}, \ and\
  \bibinfo {author} {\bibfnamefont {P.}~\bibnamefont {Podsiadlowski}},\ }\href
  {\doibase 10.1051/0004-6361/202039992} {\bibfield  {journal} {\bibinfo
  {journal} {Astron. Astrophys.}\ }\textbf {\bibinfo {volume} {650}},\ \bibinfo
  {pages} {A107} (\bibinfo {year} {2021})},\ \Eprint
  {http://arxiv.org/abs/2103.09243} {arXiv:2103.09243 [astro-ph.SR]}
  \BibitemShut {NoStop}%
\bibitem [{\citenamefont {{Han}}\ \emph {et~al.}(1995)\citenamefont {{Han}},
  \citenamefont {{Podsiadlowski}},\ and\ \citenamefont
  {{Eggleton}}}]{1995MNRAS.272..800H}%
  \BibitemOpen
  \bibfield  {author} {\bibinfo {author} {\bibfnamefont {Z.}~\bibnamefont
  {{Han}}}, \bibinfo {author} {\bibfnamefont {P.}~\bibnamefont
  {{Podsiadlowski}}}, \ and\ \bibinfo {author} {\bibfnamefont {P.~P.}\
  \bibnamefont {{Eggleton}}},\ }\href {\doibase 10.1093/mnras/272.4.800}
  {\bibfield  {journal} {\bibinfo  {journal} {\mnras}\ }\textbf {\bibinfo
  {volume} {272}},\ \bibinfo {pages} {800} (\bibinfo {year}
  {1995})}\BibitemShut {NoStop}%
\bibitem [{\citenamefont {Marchant}\ \emph {et~al.}(2018)\citenamefont
  {Marchant}, \citenamefont {Renzo}, \citenamefont {Farmer}, \citenamefont
  {Pappas}, \citenamefont {Taam}, \citenamefont {de~Mink},\ and\ \citenamefont
  {Kalogera}}]{Marchant:2018kun}%
  \BibitemOpen
  \bibfield  {author} {\bibinfo {author} {\bibfnamefont {P.}~\bibnamefont
  {Marchant}}, \bibinfo {author} {\bibfnamefont {M.}~\bibnamefont {Renzo}},
  \bibinfo {author} {\bibfnamefont {R.}~\bibnamefont {Farmer}}, \bibinfo
  {author} {\bibfnamefont {K.~M.}\ \bibnamefont {Pappas}}, \bibinfo {author}
  {\bibfnamefont {R.~E.}\ \bibnamefont {Taam}}, \bibinfo {author}
  {\bibfnamefont {S.}~\bibnamefont {de~Mink}}, \ and\ \bibinfo {author}
  {\bibfnamefont {V.}~\bibnamefont {Kalogera}},\ }\href {\doibase
  10.3847/1538-4357/ab3426} {\  (\bibinfo {year} {2018}),\
  10.3847/1538-4357/ab3426},\ \Eprint {http://arxiv.org/abs/1810.13412}
  {arXiv:1810.13412 [astro-ph.HE]} \BibitemShut {NoStop}%
\bibitem [{\citenamefont {Farmer}\ \emph {et~al.}(2019)\citenamefont {Farmer},
  \citenamefont {Renzo}, \citenamefont {de~Mink}, \citenamefont {Marchant},\
  and\ \citenamefont {Justham}}]{Farmer:2019jed}%
  \BibitemOpen
  \bibfield  {author} {\bibinfo {author} {\bibfnamefont {R.}~\bibnamefont
  {Farmer}}, \bibinfo {author} {\bibfnamefont {M.}~\bibnamefont {Renzo}},
  \bibinfo {author} {\bibfnamefont {S.}~\bibnamefont {de~Mink}}, \bibinfo
  {author} {\bibfnamefont {P.}~\bibnamefont {Marchant}}, \ and\ \bibinfo
  {author} {\bibfnamefont {S.}~\bibnamefont {Justham}},\ }\href {\doibase
  10.3847/1538-4357/ab518b} {\  (\bibinfo {year} {2019}),\
  10.3847/1538-4357/ab518b},\ \Eprint {http://arxiv.org/abs/1910.12874}
  {arXiv:1910.12874 [astro-ph.SR]} \BibitemShut {NoStop}%
\bibitem [{\citenamefont {Croon}\ \emph
  {et~al.}(2020{\natexlab{b}})\citenamefont {Croon}, \citenamefont
  {McDermott},\ and\ \citenamefont {Sakstein}}]{Croon:2020ehi}%
  \BibitemOpen
  \bibfield  {author} {\bibinfo {author} {\bibfnamefont {D.}~\bibnamefont
  {Croon}}, \bibinfo {author} {\bibfnamefont {S.~D.}\ \bibnamefont
  {McDermott}}, \ and\ \bibinfo {author} {\bibfnamefont {J.}~\bibnamefont
  {Sakstein}},\ }\href@noop {} {\  (\bibinfo {year} {2020}{\natexlab{b}})},\
  \Eprint {http://arxiv.org/abs/2007.00650} {arXiv:2007.00650 [hep-ph]}
  \BibitemShut {NoStop}%
\bibitem [{\citenamefont {{Peters}}(1964)}]{1964PhRv..136.1224P}%
  \BibitemOpen
  \bibfield  {author} {\bibinfo {author} {\bibfnamefont {P.~C.}\ \bibnamefont
  {{Peters}}},\ }\href {\doibase 10.1103/PhysRev.136.B1224} {\bibfield
  {journal} {\bibinfo  {journal} {Physical Review}\ }\textbf {\bibinfo {volume}
  {136}},\ \bibinfo {pages} {1224} (\bibinfo {year} {1964})}\BibitemShut
  {NoStop}%
\bibitem [{\citenamefont {Frieman}\ \emph {et~al.}(1987)\citenamefont
  {Frieman}, \citenamefont {Dimopoulos},\ and\ \citenamefont
  {Turner}}]{frieman1987axions}%
  \BibitemOpen
  \bibfield  {author} {\bibinfo {author} {\bibfnamefont {J.~A.}\ \bibnamefont
  {Frieman}}, \bibinfo {author} {\bibfnamefont {S.}~\bibnamefont {Dimopoulos}},
  \ and\ \bibinfo {author} {\bibfnamefont {M.~S.}\ \bibnamefont {Turner}},\
  }\href@noop {} {\bibfield  {journal} {\bibinfo  {journal} {Physical Review
  D}\ }\textbf {\bibinfo {volume} {36}},\ \bibinfo {pages} {2201} (\bibinfo
  {year} {1987})}\BibitemShut {NoStop}%
\bibitem [{\citenamefont {Raffelt}(1996)}]{raffelt1996stars}%
  \BibitemOpen
  \bibfield  {author} {\bibinfo {author} {\bibfnamefont {G.~G.}\ \bibnamefont
  {Raffelt}},\ }\href@noop {} {\emph {\bibinfo {title} {Stars as laboratories
  for fundamental physics: The astrophysics of neutrinos, axions, and other
  weakly interacting particles}}}\ (\bibinfo  {publisher} {University of
  Chicago press},\ \bibinfo {year} {1996})\BibitemShut {NoStop}%
\bibitem [{\citenamefont {Peled}\ and\ \citenamefont
  {Volansky}(2022)}]{Peled:2022byr}%
  \BibitemOpen
  \bibfield  {author} {\bibinfo {author} {\bibfnamefont {G.}~\bibnamefont
  {Peled}}\ and\ \bibinfo {author} {\bibfnamefont {T.}~\bibnamefont
  {Volansky}},\ }\href@noop {} {\  (\bibinfo {year} {2022})},\ \Eprint
  {http://arxiv.org/abs/2203.09522} {arXiv:2203.09522 [hep-ph]} \BibitemShut
  {NoStop}%
\bibitem [{\citenamefont {Di~Lella}\ \emph {et~al.}(2000)\citenamefont
  {Di~Lella}, \citenamefont {Pilaftsis}, \citenamefont {Raffelt},\ and\
  \citenamefont {Zioutas}}]{DiLella:2000dn}%
  \BibitemOpen
  \bibfield  {author} {\bibinfo {author} {\bibfnamefont {L.}~\bibnamefont
  {Di~Lella}}, \bibinfo {author} {\bibfnamefont {A.}~\bibnamefont {Pilaftsis}},
  \bibinfo {author} {\bibfnamefont {G.}~\bibnamefont {Raffelt}}, \ and\
  \bibinfo {author} {\bibfnamefont {K.}~\bibnamefont {Zioutas}},\ }\href
  {\doibase 10.1103/PhysRevD.62.125011} {\bibfield  {journal} {\bibinfo
  {journal} {Phys. Rev. D}\ }\textbf {\bibinfo {volume} {62}},\ \bibinfo
  {pages} {125011} (\bibinfo {year} {2000})},\ \Eprint
  {http://arxiv.org/abs/hep-ph/0006327} {arXiv:hep-ph/0006327} \BibitemShut
  {NoStop}%
\bibitem [{\citenamefont {Carenza}\ \emph {et~al.}(2020)\citenamefont
  {Carenza}, \citenamefont {Straniero}, \citenamefont {D\"obrich},
  \citenamefont {Giannotti}, \citenamefont {Lucente},\ and\ \citenamefont
  {Mirizzi}}]{Carenza:2020zil}%
  \BibitemOpen
  \bibfield  {author} {\bibinfo {author} {\bibfnamefont {P.}~\bibnamefont
  {Carenza}}, \bibinfo {author} {\bibfnamefont {O.}~\bibnamefont {Straniero}},
  \bibinfo {author} {\bibfnamefont {B.}~\bibnamefont {D\"obrich}}, \bibinfo
  {author} {\bibfnamefont {M.}~\bibnamefont {Giannotti}}, \bibinfo {author}
  {\bibfnamefont {G.}~\bibnamefont {Lucente}}, \ and\ \bibinfo {author}
  {\bibfnamefont {A.}~\bibnamefont {Mirizzi}},\ }\href {\doibase
  10.1016/j.physletb.2020.135709} {\bibfield  {journal} {\bibinfo  {journal}
  {Phys. Lett. B}\ }\textbf {\bibinfo {volume} {809}},\ \bibinfo {pages}
  {135709} (\bibinfo {year} {2020})},\ \Eprint
  {http://arxiv.org/abs/2004.08399} {arXiv:2004.08399 [hep-ph]} \BibitemShut
  {NoStop}%
\bibitem [{\citenamefont {Lucente}\ \emph {et~al.}(2022)\citenamefont
  {Lucente}, \citenamefont {Straniero}, \citenamefont {Carenza}, \citenamefont
  {Giannotti},\ and\ \citenamefont {Mirizzi}}]{Lucente:2022wai}%
  \BibitemOpen
  \bibfield  {author} {\bibinfo {author} {\bibfnamefont {G.}~\bibnamefont
  {Lucente}}, \bibinfo {author} {\bibfnamefont {O.}~\bibnamefont {Straniero}},
  \bibinfo {author} {\bibfnamefont {P.}~\bibnamefont {Carenza}}, \bibinfo
  {author} {\bibfnamefont {M.}~\bibnamefont {Giannotti}}, \ and\ \bibinfo
  {author} {\bibfnamefont {A.}~\bibnamefont {Mirizzi}},\ }\href@noop {} {\
  (\bibinfo {year} {2022})},\ \Eprint {http://arxiv.org/abs/2203.01336}
  {arXiv:2203.01336 [hep-ph]} \BibitemShut {NoStop}%
\bibitem [{\citenamefont {Raffelt}(1990)}]{Raffelt:1990yz}%
  \BibitemOpen
  \bibfield  {author} {\bibinfo {author} {\bibfnamefont {G.~G.}\ \bibnamefont
  {Raffelt}},\ }\href {\doibase 10.1016/0370-1573(90)90054-6} {\bibfield
  {journal} {\bibinfo  {journal} {Phys. Rept.}\ }\textbf {\bibinfo {volume}
  {198}},\ \bibinfo {pages} {1} (\bibinfo {year} {1990})}\BibitemShut {NoStop}%
\bibitem [{\citenamefont {Choplin}\ \emph {et~al.}(2017)\citenamefont
  {Choplin}, \citenamefont {Coc}, \citenamefont {Meynet}, \citenamefont
  {Olive}, \citenamefont {Uzan},\ and\ \citenamefont
  {Vangioni}}]{Choplin:2017auq}%
  \BibitemOpen
  \bibfield  {author} {\bibinfo {author} {\bibfnamefont {A.}~\bibnamefont
  {Choplin}}, \bibinfo {author} {\bibfnamefont {A.}~\bibnamefont {Coc}},
  \bibinfo {author} {\bibfnamefont {G.}~\bibnamefont {Meynet}}, \bibinfo
  {author} {\bibfnamefont {K.~A.}\ \bibnamefont {Olive}}, \bibinfo {author}
  {\bibfnamefont {J.-P.}\ \bibnamefont {Uzan}}, \ and\ \bibinfo {author}
  {\bibfnamefont {E.}~\bibnamefont {Vangioni}},\ }\href {\doibase
  10.1051/0004-6361/201731040} {\bibfield  {journal} {\bibinfo  {journal}
  {Astron. Astrophys.}\ }\textbf {\bibinfo {volume} {605}},\ \bibinfo {pages}
  {A106} (\bibinfo {year} {2017})},\ \Eprint {http://arxiv.org/abs/1707.01244}
  {arXiv:1707.01244 [astro-ph.SR]} \BibitemShut {NoStop}%
\bibitem [{\citenamefont {Friedland}\ \emph {et~al.}(2013)\citenamefont
  {Friedland}, \citenamefont {Giannotti},\ and\ \citenamefont
  {Wise}}]{Friedland:2012hj}%
  \BibitemOpen
  \bibfield  {author} {\bibinfo {author} {\bibfnamefont {A.}~\bibnamefont
  {Friedland}}, \bibinfo {author} {\bibfnamefont {M.}~\bibnamefont
  {Giannotti}}, \ and\ \bibinfo {author} {\bibfnamefont {M.}~\bibnamefont
  {Wise}},\ }\href {\doibase 10.1103/PhysRevLett.110.061101} {\bibfield
  {journal} {\bibinfo  {journal} {Phys. Rev. Lett.}\ }\textbf {\bibinfo
  {volume} {110}},\ \bibinfo {pages} {061101} (\bibinfo {year} {2013})},\
  \Eprint {http://arxiv.org/abs/1210.1271} {arXiv:1210.1271 [hep-ph]}
  \BibitemShut {NoStop}%
\bibitem [{\citenamefont {Croon}\ and\ \citenamefont
  {Sakstein}(2022)}]{djuna_croon_2022_6949679}%
  \BibitemOpen
  \bibfield  {author} {\bibinfo {author} {\bibfnamefont {D.}~\bibnamefont
  {Croon}}\ and\ \bibinfo {author} {\bibfnamefont {J.}~\bibnamefont
  {Sakstein}},\ }\href {\doibase 10.5281/zenodo.6949679} {\enquote {\bibinfo
  {title} {{Light Axion Emission and the Formation of Merging Black Holes
  Binaries}},}\ } (\bibinfo {year} {2022})\BibitemShut {NoStop}%
\bibitem [{\citenamefont {Raffelt}\ and\ \citenamefont
  {Weiss}(1995)}]{Raffelt:1994ry}%
  \BibitemOpen
  \bibfield  {author} {\bibinfo {author} {\bibfnamefont {G.}~\bibnamefont
  {Raffelt}}\ and\ \bibinfo {author} {\bibfnamefont {A.}~\bibnamefont
  {Weiss}},\ }\href {\doibase 10.1103/PhysRevD.51.1495} {\bibfield  {journal}
  {\bibinfo  {journal} {Phys. Rev. D}\ }\textbf {\bibinfo {volume} {51}},\
  \bibinfo {pages} {1495} (\bibinfo {year} {1995})},\ \Eprint
  {http://arxiv.org/abs/hep-ph/9410205} {arXiv:hep-ph/9410205} \BibitemShut
  {NoStop}%
\bibitem [{\citenamefont {Wilson}\ and\ \citenamefont
  {Nordhaus}(2022)}]{Wilson:2022elt}%
  \BibitemOpen
  \bibfield  {author} {\bibinfo {author} {\bibfnamefont {E.~C.}\ \bibnamefont
  {Wilson}}\ and\ \bibinfo {author} {\bibfnamefont {J.}~\bibnamefont
  {Nordhaus}},\ }\href@noop {} {\  (\bibinfo {year} {2022})},\ \Eprint
  {http://arxiv.org/abs/2203.06091} {arXiv:2203.06091 [astro-ph.SR]}
  \BibitemShut {NoStop}%
\bibitem [{\citenamefont {Anastassopoulos}\ \emph {et~al.}(2017)\citenamefont
  {Anastassopoulos} \emph {et~al.}}]{CAST:2017uph}%
  \BibitemOpen
  \bibfield  {author} {\bibinfo {author} {\bibfnamefont {V.}~\bibnamefont
  {Anastassopoulos}} \emph {et~al.} (\bibinfo {collaboration} {CAST}),\ }\href
  {\doibase 10.1038/nphys4109} {\bibfield  {journal} {\bibinfo  {journal}
  {Nature Phys.}\ }\textbf {\bibinfo {volume} {13}},\ \bibinfo {pages} {584}
  (\bibinfo {year} {2017})},\ \Eprint {http://arxiv.org/abs/1705.02290}
  {arXiv:1705.02290 [hep-ex]} \BibitemShut {NoStop}%
\bibitem [{\citenamefont {Barth}\ \emph {et~al.}(2013)\citenamefont {Barth}
  \emph {et~al.}}]{Barth:2013sma}%
  \BibitemOpen
  \bibfield  {author} {\bibinfo {author} {\bibfnamefont {K.}~\bibnamefont
  {Barth}} \emph {et~al.},\ }\href {\doibase 10.1088/1475-7516/2013/05/010}
  {\bibfield  {journal} {\bibinfo  {journal} {JCAP}\ }\textbf {\bibinfo
  {volume} {05}},\ \bibinfo {pages} {010} (\bibinfo {year} {2013})},\ \Eprint
  {http://arxiv.org/abs/1302.6283} {arXiv:1302.6283 [astro-ph.SR]} \BibitemShut
  {NoStop}%
\bibitem [{\citenamefont {Raffelt}(2008)}]{Raffelt:2006cw}%
  \BibitemOpen
  \bibfield  {author} {\bibinfo {author} {\bibfnamefont {G.~G.}\ \bibnamefont
  {Raffelt}},\ }\href {\doibase 10.1007/978-3-540-73518-2_3} {\bibfield
  {journal} {\bibinfo  {journal} {Lect. Notes Phys.}\ }\textbf {\bibinfo
  {volume} {741}},\ \bibinfo {pages} {51} (\bibinfo {year} {2008})},\ \Eprint
  {http://arxiv.org/abs/hep-ph/0611350} {arXiv:hep-ph/0611350} \BibitemShut
  {NoStop}%
\bibitem [{\citenamefont {Ayala}\ \emph {et~al.}(2014)\citenamefont {Ayala},
  \citenamefont {Dom\'\i{}nguez}, \citenamefont {Giannotti}, \citenamefont
  {Mirizzi},\ and\ \citenamefont {Straniero}}]{Ayala:2014pea}%
  \BibitemOpen
  \bibfield  {author} {\bibinfo {author} {\bibfnamefont {A.}~\bibnamefont
  {Ayala}}, \bibinfo {author} {\bibfnamefont {I.}~\bibnamefont
  {Dom\'\i{}nguez}}, \bibinfo {author} {\bibfnamefont {M.}~\bibnamefont
  {Giannotti}}, \bibinfo {author} {\bibfnamefont {A.}~\bibnamefont {Mirizzi}},
  \ and\ \bibinfo {author} {\bibfnamefont {O.}~\bibnamefont {Straniero}},\
  }\href {\doibase 10.1103/PhysRevLett.113.191302} {\bibfield  {journal}
  {\bibinfo  {journal} {Phys. Rev. Lett.}\ }\textbf {\bibinfo {volume} {113}},\
  \bibinfo {pages} {191302} (\bibinfo {year} {2014})},\ \Eprint
  {http://arxiv.org/abs/1406.6053} {arXiv:1406.6053 [astro-ph.SR]} \BibitemShut
  {NoStop}%
\bibitem [{\citenamefont {Viaux}\ \emph {et~al.}(2013)\citenamefont {Viaux},
  \citenamefont {Catelan}, \citenamefont {Stetson}, \citenamefont {Raffelt},
  \citenamefont {Redondo}, \citenamefont {Valcarce},\ and\ \citenamefont
  {Weiss}}]{Viaux:2013lha}%
  \BibitemOpen
  \bibfield  {author} {\bibinfo {author} {\bibfnamefont {N.}~\bibnamefont
  {Viaux}}, \bibinfo {author} {\bibfnamefont {M.}~\bibnamefont {Catelan}},
  \bibinfo {author} {\bibfnamefont {P.~B.}\ \bibnamefont {Stetson}}, \bibinfo
  {author} {\bibfnamefont {G.}~\bibnamefont {Raffelt}}, \bibinfo {author}
  {\bibfnamefont {J.}~\bibnamefont {Redondo}}, \bibinfo {author} {\bibfnamefont
  {A.~A.~R.}\ \bibnamefont {Valcarce}}, \ and\ \bibinfo {author} {\bibfnamefont
  {A.}~\bibnamefont {Weiss}},\ }\href {\doibase 10.1103/PhysRevLett.111.231301}
  {\bibfield  {journal} {\bibinfo  {journal} {Phys. Rev. Lett.}\ }\textbf
  {\bibinfo {volume} {111}},\ \bibinfo {pages} {231301} (\bibinfo {year}
  {2013})},\ \Eprint {http://arxiv.org/abs/1311.1669} {arXiv:1311.1669
  [astro-ph.SR]} \BibitemShut {NoStop}%
\bibitem [{\citenamefont {Straniero}\ \emph {et~al.}(2018)\citenamefont
  {Straniero}, \citenamefont {Dominguez}, \citenamefont {Giannotti},\ and\
  \citenamefont {Mirizzi}}]{Straniero:2018fbv}%
  \BibitemOpen
  \bibfield  {author} {\bibinfo {author} {\bibfnamefont {O.}~\bibnamefont
  {Straniero}}, \bibinfo {author} {\bibfnamefont {I.}~\bibnamefont
  {Dominguez}}, \bibinfo {author} {\bibfnamefont {M.}~\bibnamefont
  {Giannotti}}, \ and\ \bibinfo {author} {\bibfnamefont {A.}~\bibnamefont
  {Mirizzi}},\ }in\ \href {\doibase 10.3204/DESY-PROC-2017-02/straniero_oscar}
  {\emph {\bibinfo {booktitle} {{13th Patras Workshop on Axions, WIMPs and
  WISPs}}}}\ (\bibinfo {year} {2018})\ pp.\ \bibinfo {pages} {172--176},\
  \Eprint {http://arxiv.org/abs/1802.10357} {arXiv:1802.10357 [astro-ph.SR]}
  \BibitemShut {NoStop}%
\bibitem [{\citenamefont {D\'\i{}az}\ \emph {et~al.}(2019)\citenamefont
  {D\'\i{}az}, \citenamefont {Schr\"oder}, \citenamefont {Zuber}, \citenamefont
  {Jack},\ and\ \citenamefont {Barrios}}]{Diaz:2019kim}%
  \BibitemOpen
  \bibfield  {author} {\bibinfo {author} {\bibfnamefont {S.~A.}\ \bibnamefont
  {D\'\i{}az}}, \bibinfo {author} {\bibfnamefont {K.-P.}\ \bibnamefont
  {Schr\"oder}}, \bibinfo {author} {\bibfnamefont {K.}~\bibnamefont {Zuber}},
  \bibinfo {author} {\bibfnamefont {D.}~\bibnamefont {Jack}}, \ and\ \bibinfo
  {author} {\bibfnamefont {E.~E.~B.}\ \bibnamefont {Barrios}},\ }\href@noop {}
  {\  (\bibinfo {year} {2019})},\ \Eprint {http://arxiv.org/abs/1910.10568}
  {arXiv:1910.10568 [astro-ph.SR]} \BibitemShut {NoStop}%
\bibitem [{\citenamefont {Capozzi}\ and\ \citenamefont
  {Raffelt}(2020)}]{Capozzi:2020cbu}%
  \BibitemOpen
  \bibfield  {author} {\bibinfo {author} {\bibfnamefont {F.}~\bibnamefont
  {Capozzi}}\ and\ \bibinfo {author} {\bibfnamefont {G.}~\bibnamefont
  {Raffelt}},\ }\href {\doibase 10.1103/PhysRevD.102.083007} {\bibfield
  {journal} {\bibinfo  {journal} {Phys. Rev. D}\ }\textbf {\bibinfo {volume}
  {102}},\ \bibinfo {pages} {083007} (\bibinfo {year} {2020})},\ \Eprint
  {http://arxiv.org/abs/2007.03694} {arXiv:2007.03694 [astro-ph.SR]}
  \BibitemShut {NoStop}%
\bibitem [{\citenamefont {Dennis}\ and\ \citenamefont
  {Sakstein}(2023{\natexlab{a}})}]{Dennis:2023kfe}%
  \BibitemOpen
  \bibfield  {author} {\bibinfo {author} {\bibfnamefont {M.~T.}\ \bibnamefont
  {Dennis}}\ and\ \bibinfo {author} {\bibfnamefont {J.}~\bibnamefont
  {Sakstein}},\ }\href@noop {} {\  (\bibinfo {year} {2023}{\natexlab{a}})},\
  \Eprint {http://arxiv.org/abs/2305.03113} {arXiv:2305.03113 [hep-ph]}
  \BibitemShut {NoStop}%
\bibitem [{\citenamefont {Dennis}\ and\ \citenamefont
  {Sakstein}(2023{\natexlab{b}})}]{Dennis:2023ldw}%
  \BibitemOpen
  \bibfield  {author} {\bibinfo {author} {\bibfnamefont {M.}~\bibnamefont
  {Dennis}}\ and\ \bibinfo {author} {\bibfnamefont {J.}~\bibnamefont
  {Sakstein}},\ }\href@noop {} {\  (\bibinfo {year} {2023}{\natexlab{b}})},\
  \Eprint {http://arxiv.org/abs/2303.12069} {arXiv:2303.12069 [astro-ph.GA]}
  \BibitemShut {NoStop}%
\bibitem [{\citenamefont {Sakstein}\ \emph {et~al.}(2022)\citenamefont
  {Sakstein}, \citenamefont {Croon},\ and\ \citenamefont
  {McDermott}}]{Sakstein:2022tby}%
  \BibitemOpen
  \bibfield  {author} {\bibinfo {author} {\bibfnamefont {J.}~\bibnamefont
  {Sakstein}}, \bibinfo {author} {\bibfnamefont {D.}~\bibnamefont {Croon}}, \
  and\ \bibinfo {author} {\bibfnamefont {S.~D.}\ \bibnamefont {McDermott}},\
  }\href {\doibase 10.1103/PhysRevD.105.095038} {\bibfield  {journal} {\bibinfo
   {journal} {Phys. Rev. D}\ }\textbf {\bibinfo {volume} {105}},\ \bibinfo
  {pages} {095038} (\bibinfo {year} {2022})},\ \Eprint
  {http://arxiv.org/abs/2203.06160} {arXiv:2203.06160 [hep-ph]} \BibitemShut
  {NoStop}%
\bibitem [{\citenamefont {{Cox}}\ and\ \citenamefont
  {{Giuli}}(1968)}]{1968pss..book.....C}%
  \BibitemOpen
  \bibfield  {author} {\bibinfo {author} {\bibfnamefont {J.~P.}\ \bibnamefont
  {{Cox}}}\ and\ \bibinfo {author} {\bibfnamefont {R.~T.}\ \bibnamefont
  {{Giuli}}},\ }\href@noop {} {\emph {\bibinfo {title} {{Principles of stellar
  structure}}}}\ (\bibinfo {year} {1968})\BibitemShut {NoStop}%
\bibitem [{\citenamefont {{Langer}}\ \emph {et~al.}(1985)\citenamefont
  {{Langer}}, \citenamefont {{El Eid}},\ and\ \citenamefont
  {{Fricke}}}]{1985A&A...145..179L}%
  \BibitemOpen
  \bibfield  {author} {\bibinfo {author} {\bibfnamefont {N.}~\bibnamefont
  {{Langer}}}, \bibinfo {author} {\bibfnamefont {M.~F.}\ \bibnamefont {{El
  Eid}}}, \ and\ \bibinfo {author} {\bibfnamefont {K.~J.}\ \bibnamefont
  {{Fricke}}},\ }\href@noop {} {\bibfield  {journal} {\bibinfo  {journal}
  {\aap}\ }\textbf {\bibinfo {volume} {145}},\ \bibinfo {pages} {179} (\bibinfo
  {year} {1985})}\BibitemShut {NoStop}%
\bibitem [{\citenamefont {Brott}\ \emph {et~al.}(2011)\citenamefont {Brott},
  \citenamefont {de~Mink}, \citenamefont {Cantiello}, \citenamefont {Langer},
  \citenamefont {de~Koter}, \citenamefont {Evans}, \citenamefont {Hunter},
  \citenamefont {Trundle},\ and\ \citenamefont {Vink}}]{Brott:2011ni}%
  \BibitemOpen
  \bibfield  {author} {\bibinfo {author} {\bibfnamefont {I.}~\bibnamefont
  {Brott}}, \bibinfo {author} {\bibfnamefont {S.~E.}\ \bibnamefont {de~Mink}},
  \bibinfo {author} {\bibfnamefont {M.}~\bibnamefont {Cantiello}}, \bibinfo
  {author} {\bibfnamefont {N.}~\bibnamefont {Langer}}, \bibinfo {author}
  {\bibfnamefont {A.}~\bibnamefont {de~Koter}}, \bibinfo {author}
  {\bibfnamefont {C.~J.}\ \bibnamefont {Evans}}, \bibinfo {author}
  {\bibfnamefont {I.}~\bibnamefont {Hunter}}, \bibinfo {author} {\bibfnamefont
  {C.}~\bibnamefont {Trundle}}, \ and\ \bibinfo {author} {\bibfnamefont
  {J.~S.}\ \bibnamefont {Vink}},\ }\href {\doibase 10.1051/0004-6361/201016113}
  {\bibfield  {journal} {\bibinfo  {journal} {Astron. Astrophys.}\ }\textbf
  {\bibinfo {volume} {530}},\ \bibinfo {pages} {A115} (\bibinfo {year}
  {2011})},\ \Eprint {http://arxiv.org/abs/1102.0530} {arXiv:1102.0530
  [astro-ph.SR]} \BibitemShut {NoStop}%
\bibitem [{\citenamefont {Angulo}\ \emph {et~al.}(1999)\citenamefont {Angulo}
  \emph {et~al.}}]{Angulo:1999zz}%
  \BibitemOpen
  \bibfield  {author} {\bibinfo {author} {\bibfnamefont {C.}~\bibnamefont
  {Angulo}} \emph {et~al.},\ }\href {\doibase 10.1016/S0375-9474(99)00030-5}
  {\bibfield  {journal} {\bibinfo  {journal} {Nucl. Phys. A}\ }\textbf
  {\bibinfo {volume} {656}},\ \bibinfo {pages} {3} (\bibinfo {year}
  {1999})}\BibitemShut {NoStop}%
\bibitem [{\citenamefont {{Cyburt}}\ \emph {et~al.}(2010)\citenamefont
  {{Cyburt}}, \citenamefont {{Amthor}}, \citenamefont {{Ferguson}},
  \citenamefont {{Meisel}}, \citenamefont {{Smith}}, \citenamefont {{Warren}},
  \citenamefont {{Heger}}, \citenamefont {{Hoffman}}, \citenamefont
  {{Rauscher}}, \citenamefont {{Sakharuk}}, \citenamefont {{Schatz}},
  \citenamefont {{Thielemann}},\ and\ \citenamefont
  {{Wiescher}}}]{2010ApJS..189..240C}%
  \BibitemOpen
  \bibfield  {author} {\bibinfo {author} {\bibfnamefont {R.~H.}\ \bibnamefont
  {{Cyburt}}}, \bibinfo {author} {\bibfnamefont {A.~M.}\ \bibnamefont
  {{Amthor}}}, \bibinfo {author} {\bibfnamefont {R.}~\bibnamefont
  {{Ferguson}}}, \bibinfo {author} {\bibfnamefont {Z.}~\bibnamefont
  {{Meisel}}}, \bibinfo {author} {\bibfnamefont {K.}~\bibnamefont {{Smith}}},
  \bibinfo {author} {\bibfnamefont {S.}~\bibnamefont {{Warren}}}, \bibinfo
  {author} {\bibfnamefont {A.~e.}\ \bibnamefont {{Heger}}}, \bibinfo {author}
  {\bibfnamefont {R.~D.}\ \bibnamefont {{Hoffman}}}, \bibinfo {author}
  {\bibfnamefont {T.}~\bibnamefont {{Rauscher}}}, \bibinfo {author}
  {\bibfnamefont {A.~e.}\ \bibnamefont {{Sakharuk}}}, \bibinfo {author}
  {\bibfnamefont {H.}~\bibnamefont {{Schatz}}}, \bibinfo {author}
  {\bibfnamefont {F.~K.}\ \bibnamefont {{Thielemann}}}, \ and\ \bibinfo
  {author} {\bibfnamefont {M.}~\bibnamefont {{Wiescher}}},\ }\href {\doibase
  10.1088/0067-0049/189/1/240} {\bibfield  {journal} {\bibinfo  {journal}
  {\apjs}\ }\textbf {\bibinfo {volume} {189}},\ \bibinfo {pages} {240}
  (\bibinfo {year} {2010})}\BibitemShut {NoStop}%
\end{thebibliography}%
